\documentclass[12pt]{article}
\usepackage[dvips]{epsfig}
\usepackage{cite,overcite,amsmath,amssymb,dcolumn}
\usepackage{typearea}\typearea{22}
\newcommand{\mul}[1]{\multicolumn{1}{c}{#1}}
\newcommand{\mult}[1]{\multicolumn{2}{c}{#1}}
\newcommand{\opC}{\operatorname{C}}
\newcommand{\opH}{\operatorname{H}}
\begin{document}
\title{Local Structure and Dynamics of {\it Trans}-polyisoprene oligomers}
\author{Roland Faller$^{1,}$\footnotemark \and Florian M{\"u}ller-Plathe$^1$
  \and  
  Manolis Doxastakis$^2$ \and Doros Theodorou$^2$ \\
  \small $^1$ Max-Planck-Institut f\"ur Polymerforschung, Ackermannweg 10,
  D-55128 Mainz, Germany\\
  \small $^2$ Department of Chemical Engineering, University of Patras \\ 
  \small and ICE/HT-FORTH , GR-26500 Patras, Greece
  }
\maketitle
\setcounter{footnote}{1}
\renewcommand{\thefootnote}{\fnsymbol{footnote}}
\footnotetext{Corresponding author, new address: Department of Chemical
  Engineering, University of Wisconsin, Madison, WI 53706, USA}
\abstract{
  \noindent Mono- and poly-disperse melts of oligomers (average length 10
  monomers) of {\it trans}-1,4-polyisoprene are simulated in full atomistic
  detail. The force-field is developed by means of a mixture of {\it ab
  initio} quantum-chemistry and an automatic generation of empirical
  parameters. Comparisons to NMR and scattering experiments validate the
  model. The local reorientation dynamics shows that for C$-$H vectors there is
  a two-stage process consisting of an initial decay and a late-stage
  decorrelation originating from overall reorientation. The atomistic model
  can be successfully mapped onto a simple model including only beads
  for the monomers with bond springs and bond angle potentials.
  End-bridging Monte Carlo as an equilibration stage and molecular dynamics
  as the subsequent simulation method together prove to be a useful method for
  polymer simulations. 
}
\clearpage
\section{Introduction}
The understanding of polymer materials from the very local to the macroscopic
scale is at the focus of theoretical material science. Simulations are a
useful means for reaching this goal.  Fully detailed simulations incorporating
interaction centers for all atoms allow extensive investigations of
polymer-specific models to compare directly to high resolution experiments
like neutron scattering\cite{moe99}, positronium annihilation
spectroscopy\cite{schmitz00a} or nuclear magnetic 
resonance\cite{mplathe00}. In contrast to simple generic models\cite{grest86},
atomistic simulations capture the differences between polymer species. This
understanding is a prerequisite for the design of tailormade materials for
special applications.

The present contribution investigates, as an example, oligomers of {\it
trans}-1,4-polyisoprene.  We focus on {\it local} structure and dynamics as
the local scale is influenced mainly by the specific chemical structure. In
this range all-atom models are necessary for a reliable static and dynamic
investigation\cite{gunsteren90}. Moreover we compare our results to NMR
measurements, which always sample atom-atom vectors. Therefore, it is useful
for the analysis to have all atoms present.  For the properties under
investigation short chains are sufficient, as the connectivity and
non-crossability contribute to phenomena on the generic large-scale
level\cite{kremer90,faller00sa} which is not of interest here.  Several
experiments have been performed aimed at the local properties of polyisoprene
like reorientation times and the
packing\cite{batie89,denault89,zorn92,laupretre93,dlubek98,english98}.
Additionally, simulations of pure {\it cis}-polyisoprene are already
known\cite{moe96a,moe99}. To our knowledge there are, however, not yet
simulations of the {\it trans} conformer. Industrially this conformer is not
as important as the {\it cis}-conformer. However, typical polyisoprene
materials include several percent {\it trans} content. In order to understand
the influence of this content, it is worth investigating pure {\it
trans}-polyisoprene and look for differences from the other conformers.
\section{The simulation model and technical details}
\label{sec:ff}
\subsection{Starting structures from quantum chemistry}
Three different chain length mixtures, each containing 100 oligomers (13200
atoms) of polyisoprene with an average molecular weight of 682~g/mol
(corresponding to 10 monomers) are simulated at temperatures of 300~K and
413~K. One system is mono-disperse (in the following referred to as system~1)
and is simulated at constant density of 890~kg/m$^3$ (experimental density of 
the {\it cis}-conformer: 900~kg/m$^3$)\cite{fetters99a} in an orthorhombic box
under periodic boundary conditions with box lengths:
$4.97\text{nm}\times5.16\text{nm}\times4.97\text{nm}$. This is about two and
a half times the end-to-end distance of the chains, thus artifacts from 
interactions of chains with their mirror images are absent. The torsional 
conformations were set up in the equilibrium distribution derived from quantum
chemical calculations. Twenty-seven configurations of a dimer were 
investigated with quantum chemical calculations (for details see 
Section~\ref{sec:ffex}). Only seven of them are actually at low energy 
(Table~\ref{tab:relmin}) and only their statistics was used to generate initial
torsional distributions. The chains were packed in a simulation box of
$9\text{nm}\times9\text{nm}\times9\text{nm}$ and compressed to the final
density in steps, using molecular dynamics. During this procedure anisotropic 
box fluctuations were allowed to better equilibrate stresses in the box 
leading to the orthorhombic box explained above. At the same time the time-step
was increased from 0.01~fs to 1~fs.
\subsection{Starting structure via End-bridging Monte Carlo}
The two polydisperse systems (systems~2 and~3) were additionally equilibrated
using the end-bridging Monte Carlo (EBMC) procedure\cite{pant95,mavrantzas99}
before the molecular dynamics simulations were performed at a constant
pressure of 101.3~kPa to determine the density in a poly-disperse melt.
We used two systems with slightly different chain length distributions (see 
below) in order to increase the statistics and to look for influences of the
actual realization of the ensemble.

The EBMC was performed in a united-atom model as the positions of the 
hydrogens are not important at this stage. The non-bonded interaction 
potential parameters for this procedure are shown in Table~\ref{tab:ebff}. All
Monte Carlo steps are accepted according to the Metropolis criterion.

\begin{figure}
  \begin{center}
    \includegraphics[width=0.5\linewidth]{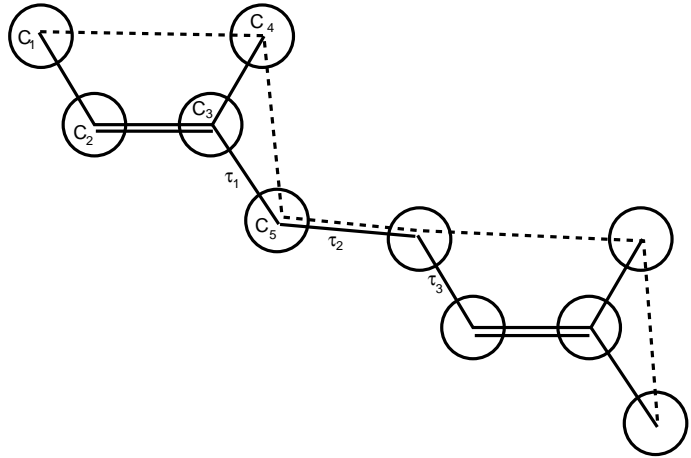}
    \caption{Carbons in the dimer of {\it trans}-polyisoprene with definition
      of the torsions. Additionally, the hypothetical chain for the
      end-bridging moves is shown as dashed line.} 
    \label{fig:PI}
  \end{center}
\end{figure}

The end-bridging steps were performed using a hypothetical chain (see dashed
line in Figure~\ref{fig:PI}). The monomers on this chain consist only of the
carbons 1, 4, and 5. Unlike the original procedure for 
polyethylene\cite{pant95,mavrantzas99}, moves can
only be performed breaking the bonds between carbon 5 and 1 as the topology of
{\it trans}-polyisoprene must not be altered. As the bridging procedure
always needs three trimers: one to bridge from, one to bridge to, and one 
as the bridge, the three carbons of one pseudo-monomer act as such trimers. 
After the respective move the positions of carbons~2 and~3 have to be 
recalculated. Since the double bond keeps
the whole monomer in plane the positions are defined exactly if bond lengths
and angles are left unchanged. In the Monte Carlo procedure no bond length or
bond angle was changed, neither for the hypothetical chain, nor for the
atomistic chain. Any of these local degrees of freedom were left for the
molecular dynamics to equilibrate.

\begin{figure}
  \begin{center}
    \includegraphics[angle=-90,width=0.5\linewidth]{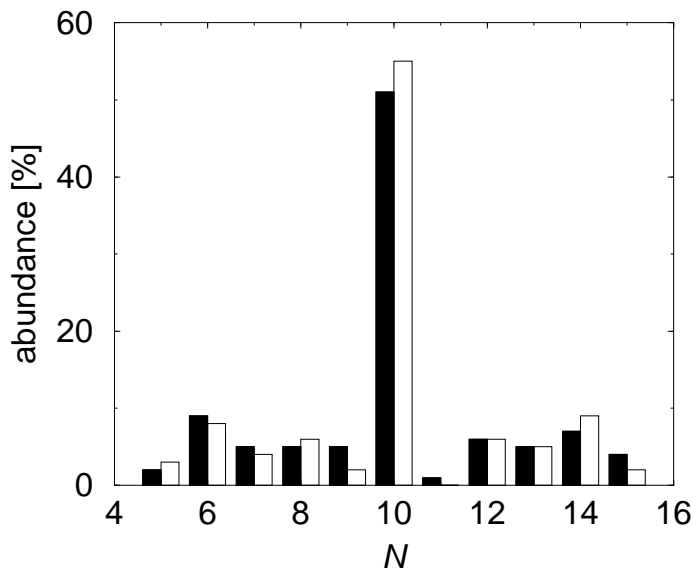}
    \caption{Molecular weight distributions of systems~2 (filled bars) and~3
      (open bars)}
    \label{fig:distMW}
  \end{center}
\end{figure}

\begin{table}
  \begin{center}
    \[
    \begin{array}{lD{.}{.}{-1}D{.}{.}{-1}}
      \hline
      \mul{\text{Interaction}} & \mul{\sigma [nm]} & 
      \mul{\epsilon[kJ/mol]}\\
      \hline
      \text{C$-$C},\text{C$-$CH},\text{CH$-$CH}  & 0.38   & 0.4186 \\
      \text{C$-$CH}_2,\text{CH$-$CH}_2           & 0.4257 & 0.4249 \\      
	\text{C$-$CH}_3,\text{CH$-$CH}_3           & 0.4257 & 0.6299 \\
      \text{CH}_2\text{$-$CH}_2                & 0.45   & 0.3918 \\
      \text{CH}_2\text{$-$CH}_3                & 0.45   & 0.6095 \\
      \text{CH}_3\text{$-$CH}_3                & 0.45   & 0.9479 \\
      \hline
    \end{array}
    \]
  \end{center}
  \caption{Nonbonded parameters (Lennard-Jones 12-6) used in the united atom
    end-bridging Monte Carlo simulations of {\it trans}-polyisoprene. Bond
    lengths, angles and torsion potentials are the same as in the all atom
    simulations. All interactions within a topological distance of up to~3
    bonds are excluded, as they are taken into account in the torsion
    potentials.}
  \label{tab:ebff}
\end{table}
Anisotropic box fluctuations were allowed but no shearing. The systems are
slightly poly-disperse, as this version of EBMC does not work at constant
topology, i.e. chain length distribution. The molecular weight distribution is
sharply peaked at $N=10$ (Figure~\ref{fig:distMW}) as we did not intend to
equilibrate the chain length distribution but to investigate nearly
mono-disperse systems. We started with a monodisperse sample and let the
end-bridging procedure run for a limited time with chemical potential $\mu=0$
for $5\le N\le15$ and $\mu=-\infty$ elsewhere. Strict monodispersity, however,
would require much longer equilibration times as EBMC would be prohibited. On
average, every chain underwent 2.8 successful end-bridging moves in addition
to many local moves before the molecular dynamics started. The resulting 
distribution shows two weak side-peaks at $N\approx6$ and $N\approx14$. These
can be explained by the procedure. Chain lengths of 9 or 11 are not directly 
accessible by endbridging moves starting from the initial monodisperse ($N=10$)
sample, as no single monomers can be transferred. Since they have to be 
reached indirectly, they have a low probability.

\subsection{The force-field}
\label{sec:ffex}
As the torsion potentials are very important for the configurations of the
chains, quantum chemical calculations were performed with the packages
Gaussian~94\cite{gaussian94} and Gaussian~98\cite{gaussian98}. The energy
differences between different minima were calculated with hybrid density
functional calculations (B3LYP)\cite{becke93} and the barrier heights are
calculated by Hartree-Fock using a 6-31G** basis set on a dimer of
trans-polyisoprene.  In order to calculate barrier heights, constrained
optimizations at fixed bond lengths are applied where the torsions are
changed in steps of 15 or 30~degrees. The results were fitted to a Fourier
series in the torsion angle with four terms. The constant energy shift is
discarded as it does not enter the forces
\begin{equation}
  V_{\text{tors}}=\sum_{i=1}^{3}k_{\tau}/2\Big[1-\cos[i(\tau-\tau_0)]\Big].
\end{equation}
The resulting parameters are shown in Table~\ref{tab:torspot}. The relevant
minima are first supposed to occur in {\it trans} and {\it gauche} states.
\begin{table}
  \begin{center}
    \begin{tabular}{crcrcc}
      \hline
      torsion & \mult{dihedral angle} & \mult{strength} & periodicity \\ 
      & \mult{$\tau_0$ [degrees]} & \mult{$k_{\tau}$[kJ/mol]} & $i$ \\
      \hline
      1 &   0 & &  5.2 & & 1 \\
      1 &   0 & & -7.4 & & 2 \\
      1 &   0 & & 10.0 & & 3 \\
      2 & 180 & &  9.7 & & 1 \\
      2 & 180 & & 14.1 & & 3 \\
      3 &   0 & &-21.1 & & 1 \\
      3 &   0 & &-12.3 & & 2 \\
      3 &   0 & &  0.5 & & 3 \\
      \hline
    \end{tabular}
    \caption{Force-field parameters for the torsion angles. For the
      nomenclature see Figure~\ref{fig:PI}.}
    \label{tab:torspot}
  \end{center}
\end{table} 
\begin{table}
  \begin{center}
    \[
    \begin{array}{cD{.}{.}{-1}D{.}{.}{-1}D{.}{.}{-1}D{.}{.}{-1}}
      \hline
      \text{Conf.} & \mul{\text{Torsion 1}} & \mul{\text{Torsion 2}} & 
      \mul{\text{Torsion 3}} & 
      \mul{\Delta E [\text{kJ/mol}]}\\
      & \mul{\text{C$_2-$C$_3-$C$_5-$C$_1$}} 
      & \mul{\text{C$_3-$C$_5-$C$_1-$C$_2$}}
      & \mul{\text{C$_5-$C$_1-$C$_2-$C$_3$}} & \\
      \hline
      1 & 180.0  & 180.0  & 180.0  & 42.56\\
      2 & -111.2 & 176.1  & 115.7  & 0.33\\
      3 & 117.4  & -62.9  & 133.5  & 2.87\\
      4 & 115.3  & -63.8  & -124.0 & 2.28\\
      5 & -109.3 & -177.7 & -115.3 & 0.00\\
      6 & -94.9  & -70.5  & 123.6  & 4.91\\
      7 & -86.0  & -59.0  & -108.3 & 6.72\\
      \hline
    \end{array}
    \]
    \caption{Relevant torsion conformations after geometry optimization using
      the hybrid method 6$-$311G**/B3LYP.} 
   \label{tab:relmin}
 \end{center}
\end{table}
The initial torsion states are set up according to this distribution (before
EBMC) in a Markov chain way, i.e. the chains are built monomer by monomer and
the triple of torsions connecting two monomers is selected by random numbers
according the energy distribution.

The bond angle potential and the improper dihedral potential, which keeps the
atoms at the double bond in plane, are harmonic with force constants adapted
from previous simulations of small molecules
(Table~\ref{tab:ffangbnd}).\cite{faller99c,schmitz99a} The harmonic dihedral
has a strength of 160~kJ/(mol*rad$^2$) and is applied to the dihedrals
C$_1-$C$_2-$C$_3-$C$_4$, H$_{\text{C2}}-$C$_2-$C$_3-$C$_5$, and
C$_2-$C$_3-$C$_4-$C$_5$. 
\begin{table}
  \[
    \begin{array}{crc|cD{.}{.}{-1}}
      \hline
      \text{angle} & \mul{\phi_{0}} & k \text{[kJ/(mol*rad}^2)] & \text{bond} 
      & l_{b} \text{[nm]} \\
      \hline
      \opC_1-\opC_2-\opC_3 & 128.7 & 250 & \opC_1-\opC_2 & 0.150 \\
      \opC_2-\opC_3-\opC_4 & 124.4 & 250 & \opC_2=\opC_3 & 0.1338 \\
      \opC_2-\opC_3-\opC_5 & 120.2 & 250 & \opC_3-\opC_4 & 0.151 \\
      \opC_4-\opC_3-\opC_5 & 115.4 & 250 & \opC_3-\opC_5 & 0.1515 \\
      \opC_3-\opC_5-\opC_1 & 114.5 & 250 & \opC_5-\opC_1 & 0.155 \\
      \opC_5-\opC_1-\opC_2 & 112.7 & 250 & \opC-\opH & 0.109 \\
      \opC-\opC_{\text{sp3}}-\opH & 109.5 & 250 & & \\
      \opC_1-\opC_2-\opH & 114.4 & 250 & & \\
      \opC_3-\opC_2-\opH & 114.4 & 250 & & \\
      \opH-\opC-\opH  & 109.5 & 250 & & \\
      \hline
    \end{array}
    \]
    \caption{Angles and bond lengths for atomistic simulations. The
      equilibrium values of the angles $\phi_{0}$ and the bond lengths $l_{b}$
      are derived from density functional calculations using
      Gaussian~94\cite{gaussian94} and experimental data\cite{CRC},  $k$ is the
      force constant for the harmonic bond angle potential. The bond lengths
      are constrained.}
    \label{tab:ffangbnd}
\end{table}
Also the non-bonded interaction parameters (Table~\ref{tab:ff}) are adapted
from simulations of cyclohexene.\cite{schmitz99a} They were derived
using the automatic simplex parameterization.\cite{faller99c} The polymer
density was not fully satisfactory with the original cyclohexene
force-field.\cite{faller00a} Thus, the $\sigma$ value of the hydrogens was
slightly increased to reproduce the experimental density at 300~K.
Lennard-Jones interactions between unlike atoms were based on the
Lorentz-Berthelot mixing rules.\cite{allen87} 
\begin{table}
  \[
    \begin{array}{cD{.}{.}{-1}D{.}{.}{-1}D{.}{.}{-1}}
      \hline
      \text{atom} & \mul{m\text{[amu]}} & \mul{\sigma\text{[nm]}} & 
      \mul{\epsilon\text{[kJ/mol]}}\\ 
      \hline
      \opC_{\text{sp2}} & 12.01 & 0.321 & 0.313\\
      \opC_{\text{sp3}} & 12.01 & 0.311 & 0.313\\
      \opH & 1.00782 & 0.24 & 0.2189\\
      \hline
    \end{array}
    \]
    \caption{Force-field parameters for the non-bonded interactions. $m$ is
      the atom mass, $\sigma$ the interaction radius, and $\epsilon$ the
      interaction strength. }
    \label{tab:ff}
\end{table}

\begin{table}
  \[
    \begin{array}{ccrrD{.}{.}{-1}}
      \hline
      \text{system} & \text{T[K]} & t_{sim}\text{[ps]} 
      & \frac{M_w}{M_n} & \mul{\rho\text{[kg/m}^3]} \\
      \hline
      1 & 300 & 1184 & 1.00 & 890\\
      2 & 300 & 2012 & 1.05 & 917.4\\
      3 & 300 & 1737 & 1.05 & 916.8\\
      3 & 413 &  792 & 1.05 & 826\\
      \hline
    \end{array}
    \]
    \caption{Simulation parameters for the three systems under study. Before
      the above-mentioned simulation time was started a few hundred
      picoseconds equilibration time was waited for.}
    \label{tab:simlen}
\end{table}

Atoms connected by any bonding potential did not interact by the Lennard-Jones
potential. Additionally, the following non-bonded interactions were
excluded: all within one monomer, and all C$-$C, C$-$H and H$-$H
interactions between the second half of the carbons of one monomer (atoms
C$_3$, C$_4$, and C$_5$) and the first half of its following neighbor (C$_1$
and C$_2$) including the hydrogens connected to them for system~1. For
system~2 and~3 the latter C$-$H and H$-$H interactions were not excluded (only
up to ``$1-4$'' interactions), which leads to differences as the different
torsion states alter the distances between these atoms so that the effective
energies of the torsions are shifted. This happened due to an oversight in the
initial simulations. As the simulations are very time consuming we deemed a 
complete repetition not necessary.

Constant temperature and pressure are ensured using Berendsen's
method~\cite{berendsen84} with time constants 0.2~ps for temperature and 8~ps
(system~2) or 12~ps (system~3) for pressure, respectively. The pressure
coupling using a compressibility of $2\times10^{-7}~\text{kPa}^{-1}$ was
employed for the three directions independently. All simulations were performed
using the YASP molecular simulation package\cite{yasp} with a time-step of 1~fs
and a cutoff for the non-bonded interactions at 0.9~nm. Configurations were
saved every picosecond. Simulation lengths and polydispersities are shown in
Table~\ref{tab:simlen}. Bond lengths are constrained to the desired values of
Table~\ref{tab:ffangbnd} using the SHAKE algorithm.\cite{ryckaert77,mplathe91}
\section{Thermodynamic and static structural properties}
\subsection{Density}
The systems simulated under constant pressure conditions arrived at densities
of 917.4~kg/mol (system 2) and 916.8~kg/mol (system 3), respectively, which is
a discrepancy of less than 2\% relative to the experimental value of
900~kg/mol for chains of 16 monomer length~\cite{fetters99a}. The isochoric
simulation of system 1 had a pressure of -500~kPa. In $NVT$ simulations, a
negative pressure of this magnitude means that the system is not exactly at
the correct density but would like to contract a little further. As all
densities are quite close to the experimental value, which itself is not too
certain, since it was determined for a mixture of {\it cis}- and {\it
trans}-PI, a closer refinement of the force-field parameters was not deemed
necessary. The density also depends weakly on the intra-chain part of the
potential. If the torsion potentials are switched off, the density increases
by about 2\%, as the chains can pack more effectively (System~1).
\subsection{Single chain properties}
The mean squared end-to-end distance of the oligomers was measured between the
terminating carbons (C$_1^{\text{mono 1}}$ and C$_5^{\text{mono~}n}$). The
respective result of 5.12~nm$^2$ corresponds to 0.0075~nm$^2$mol/g for the
monomer-weight-specific end-to-end distance.  The experimental value of
0.0060~nm$^2$mol/g per monomer is for a mixture of {\it trans} and {\it cis}
polyisoprene. The experiments are performed doing small angle neutron
scattering in a melt of 7-mers.\cite{fetters94}

\begin{figure}
  \begin{center}
    \includegraphics[angle=-90,width=0.5\linewidth]{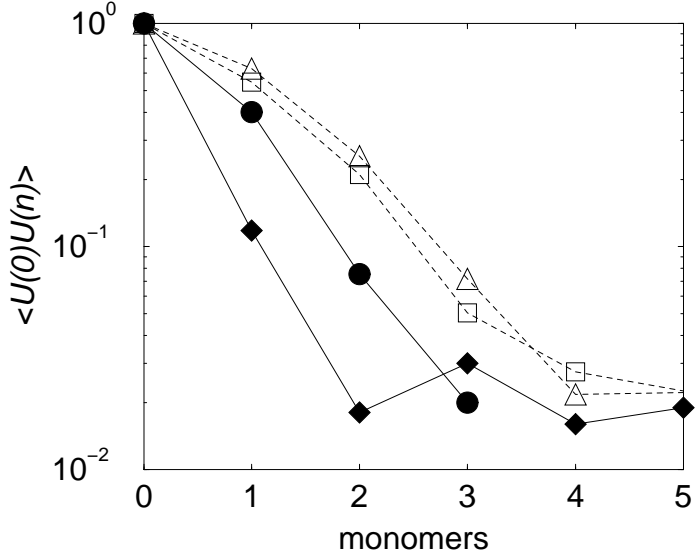}
  \end{center}
  \caption{Bond vector correlation functions in the atomistic polyisoprene
    simulations of system~1. The filled symbols are correlations between
    direction vectors in the monomers. (diamond: double-bonds, circle
    C$_1-$C$_5$) The open symbols correspond to vectors connecting atoms
    belonging to neighboring monomers (squares: C$_2$, triangles C$_1$).
    }
  \label{fig:atobc}
\end{figure}

The persistence length $l_p$ measures direction correlations of unit vectors
along the chain. It is calculated using suitably defined points along the
backbone. The function
\begin{equation}
  \big\langle\vec{u}(\vec{r})\cdot\vec{u}(\vec{r}_0)\big\rangle=
  e^{-\frac{|\vec{r}-\vec{r}_0|}{l_p}}
\end{equation}
is fitted against the bond correlation. There is some freedom in how to define
the tangent vector in an atomistic model. The vectors connecting the C$_1$ (or
the C$_2$) atoms of adjacent monomers were investigated. Additionally,
intra-monomer vectors along the double-bond and vectors connecting the two end
carbons of the same monomer (C$_1-$C$_5$) are included. The bond-correlation
functions are not really exponential (Figure~\ref{fig:atobc}), as the
interactions are complex and the chains are short. Thus, persistence lengths
deduced from the fitting procedure can only be taken as estimates.
End-effects were not excluded for reasons of statistics. The correlation
functions are different for different vectors. The vector representing the
shortest connection (the double-bond) has the shortest correlation length.
The persistence lengths range between 0.5 (for the double bonds) and 1.5 (for
the intermonomer vectors) in monomer diameters ($\approx
0.3$~nm). Polyisoprene is, therefore, rather flexible already on the length
scale of a few monomers. The monomer itself is intrinsically stiff, but the
three torsions between neighboring monomers provide a flexible link.
\subsection{Local packing of chains}
\begin{figure}
  \begin{center}
    \includegraphics[angle=-90,width=0.45\linewidth]{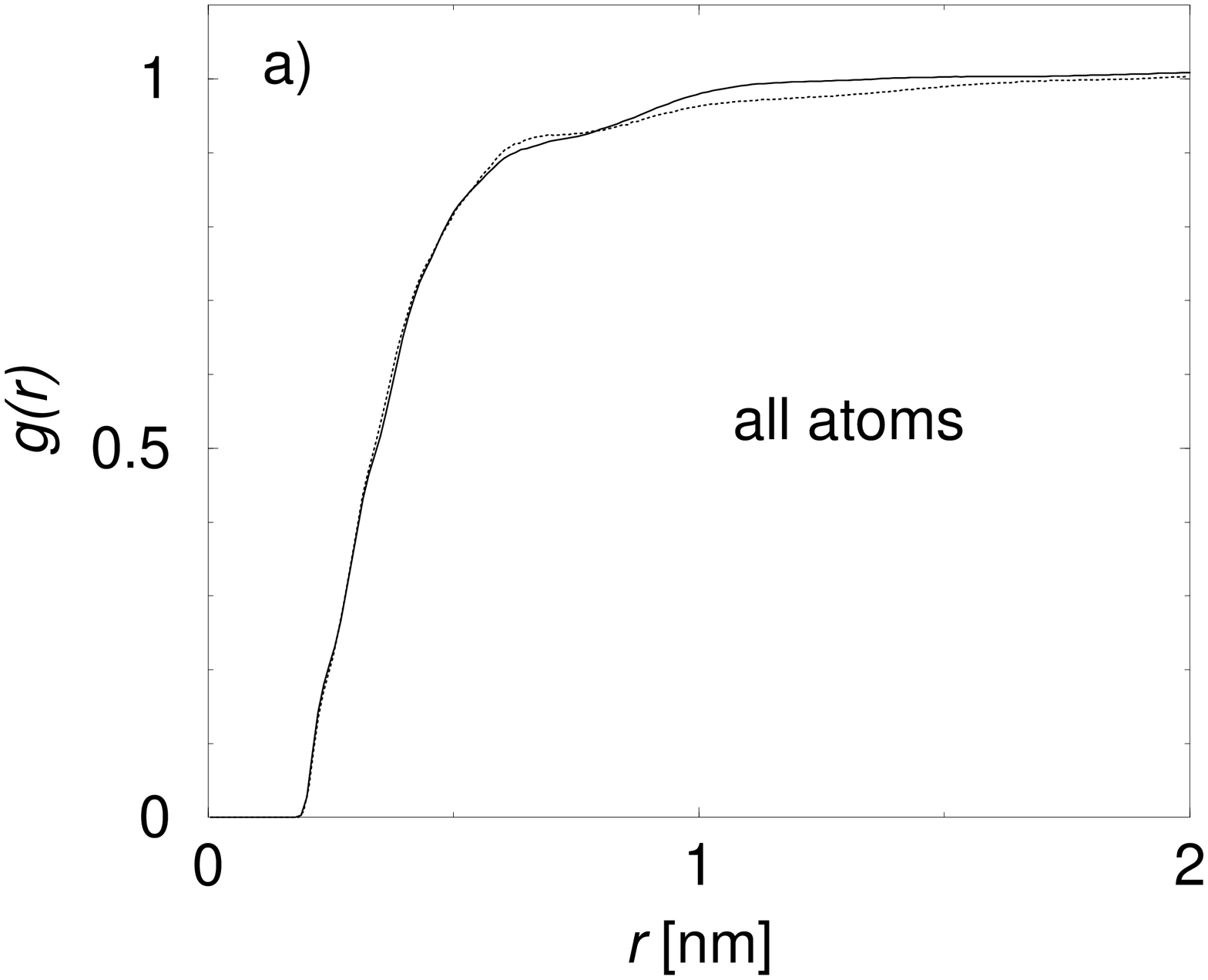}
    \includegraphics[angle=-90,width=0.45\linewidth]{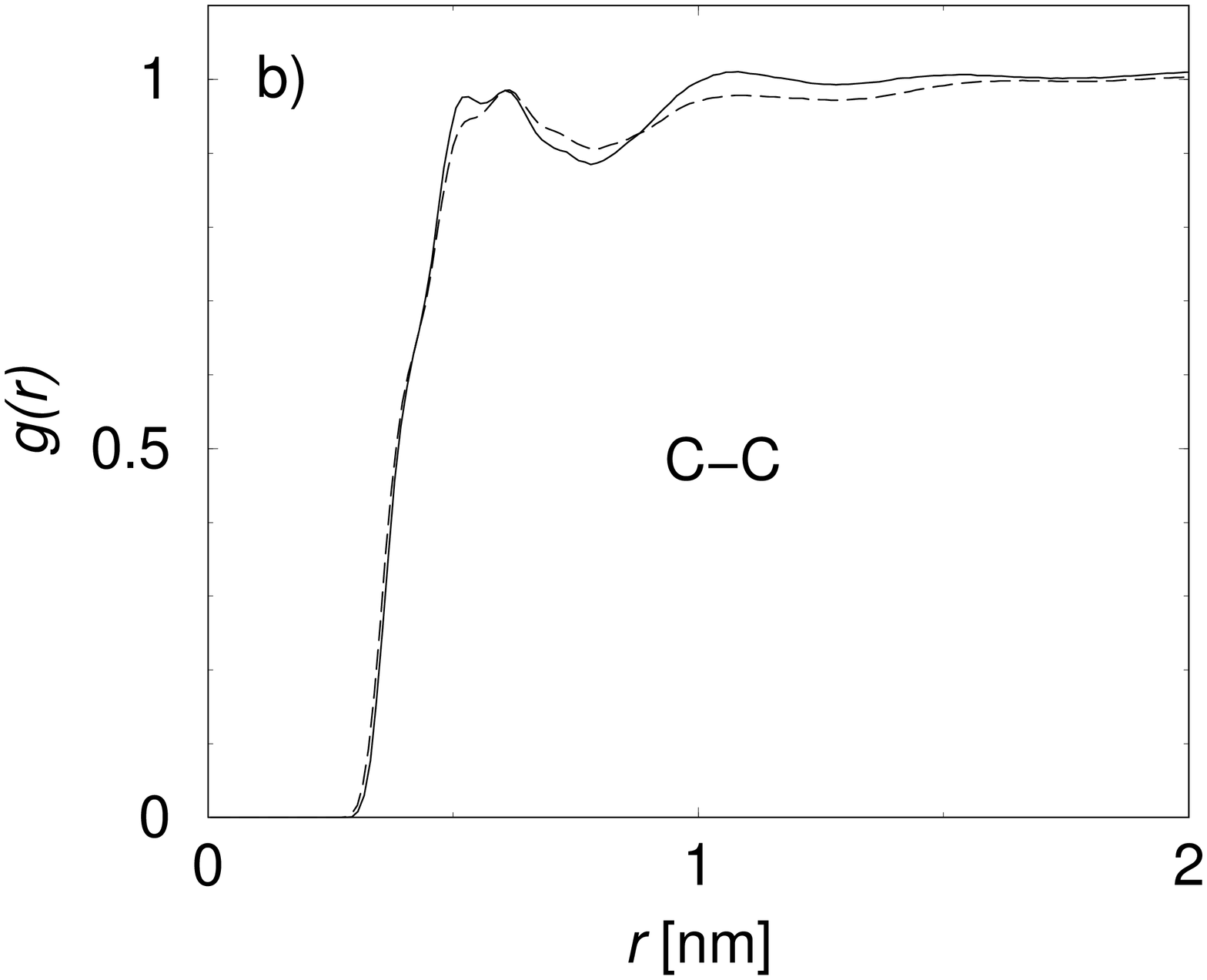}
    \includegraphics[angle=-90,width=0.45\linewidth]{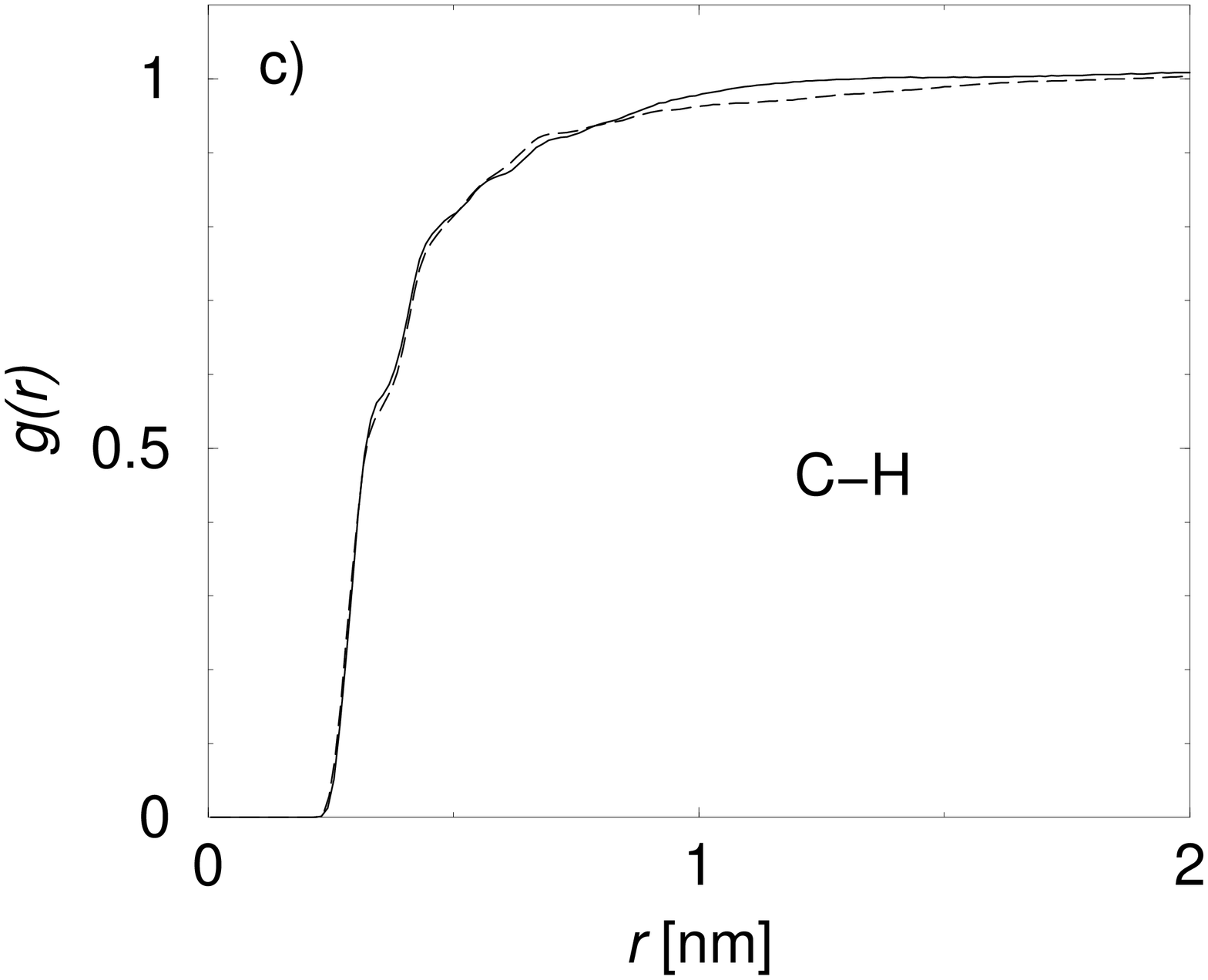}
    \includegraphics[angle=-90,width=0.45\linewidth]{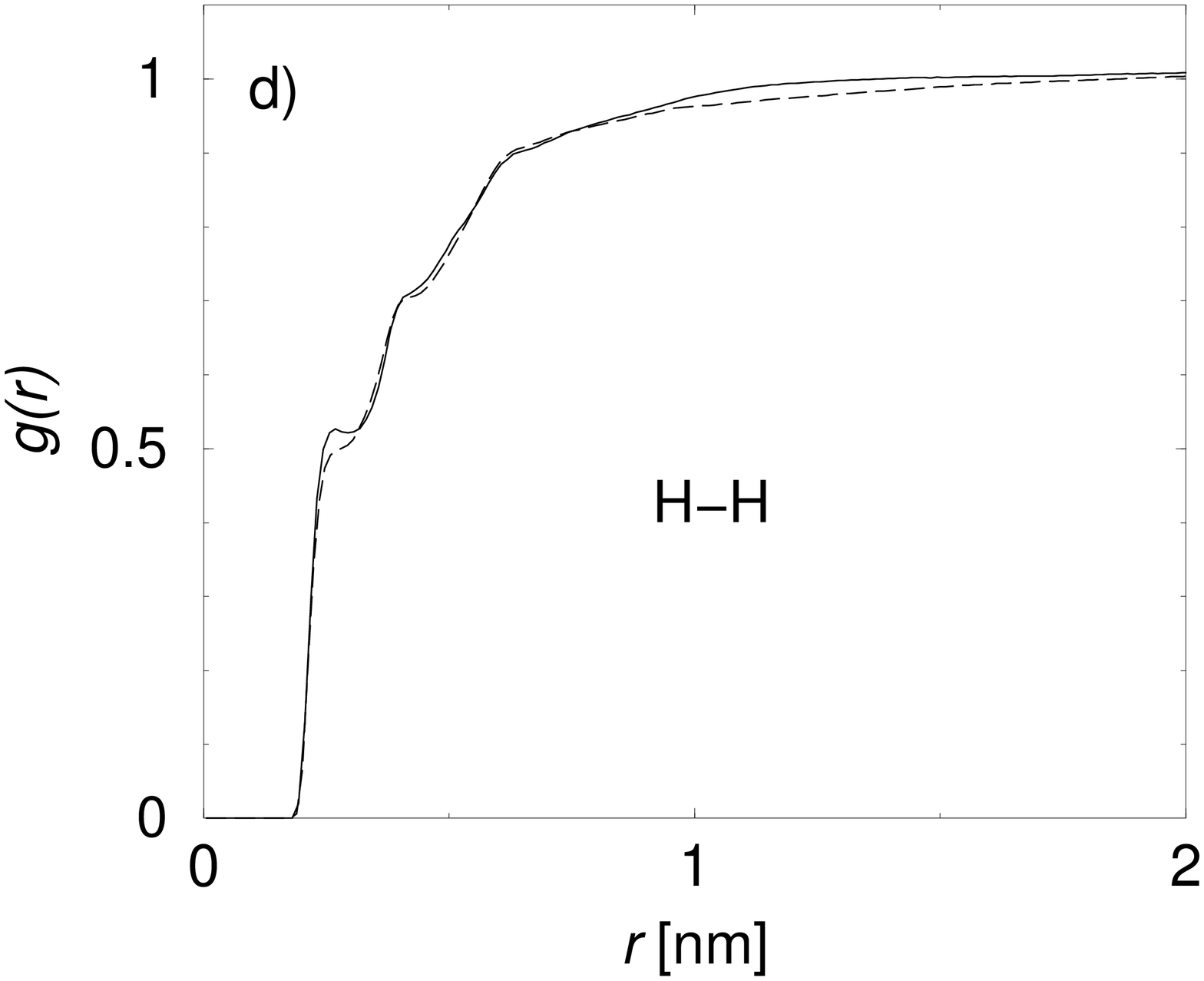}
    \caption{Interchain radial distributions of the different atomic pairs
      (full line system~1, broken line system~2) a) all atoms, b) only
      carbons, c) only carbon hydrogen d) only hydrogen pairs}
    \label{fig:rdf}
  \end{center}
\end{figure}

The local structure in the melt can be characterized by different pair
distribution functions. Figure~\ref{fig:rdf} shows inter-chain radial
distribution functions of the different atom pairs present in polyisoprene.
The curves between the different systems differ only slightly due to density 
differences. The conspicuous absence of a distinctive first peak in the 
all-atom and the hydrogen $g(r)$ at short distances indicates that chains 
cannot easily interpenetrate. Structure is averaged out in the all-atom $g(r)$,
in contrast to, e.g., the carbon-carbon distribution function, which exhibits 
several clear peaks. However, also these do not rise to values 
$g_{\text{CC}}(r)>1$.

\begin{figure}
  \[
  \includegraphics[angle=-90,width=0.5\linewidth]{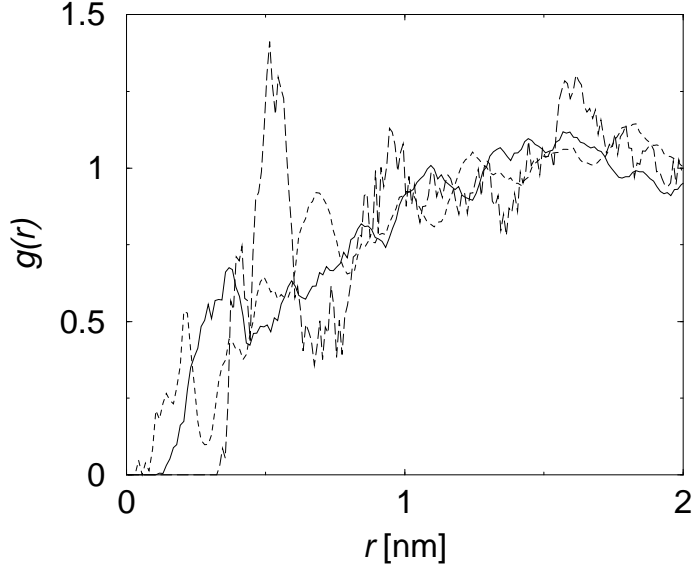}
  \]
  \caption{Center of mass radial distribution function of the atomistic
    polyisoprene chains. A correlation hole on local scales is seen, on bigger
    length scales the distribution is flat. The dashed line corresponds
    to system~1, the solid line to system~2, the dotted line to system~3.} 
  \label{fig:chainchainrdf}
\end{figure}

Figure~\ref{fig:chainchainrdf} shows the center-of-mass distribution of whole
chains.
In spite of the statistical noise it is clear that they can approach as close
as 0.2~nm. The similarity of the RDFs for systems~2 and~3 suggests that the
combination of EBMC and MD leads to a reliable structure. At distances of
more than 1~nm the distribution is quite flat, at shorter distances a 
correlation hole is
clearly visible. System~1, which was not equilibrated using the end-bridging,
exhibits unrealistically sharp peaks. The differences in the molecular weight 
distribution also contribute to the different structures. However, molecular 
dynamics alone is not able to equilibrate a simulation of this size. The 
introduction of EBMC brings us a good step further, although some remnants of 
the setup may still be present.

\begin{figure}
  \includegraphics[angle=-90,width=0.5\linewidth]{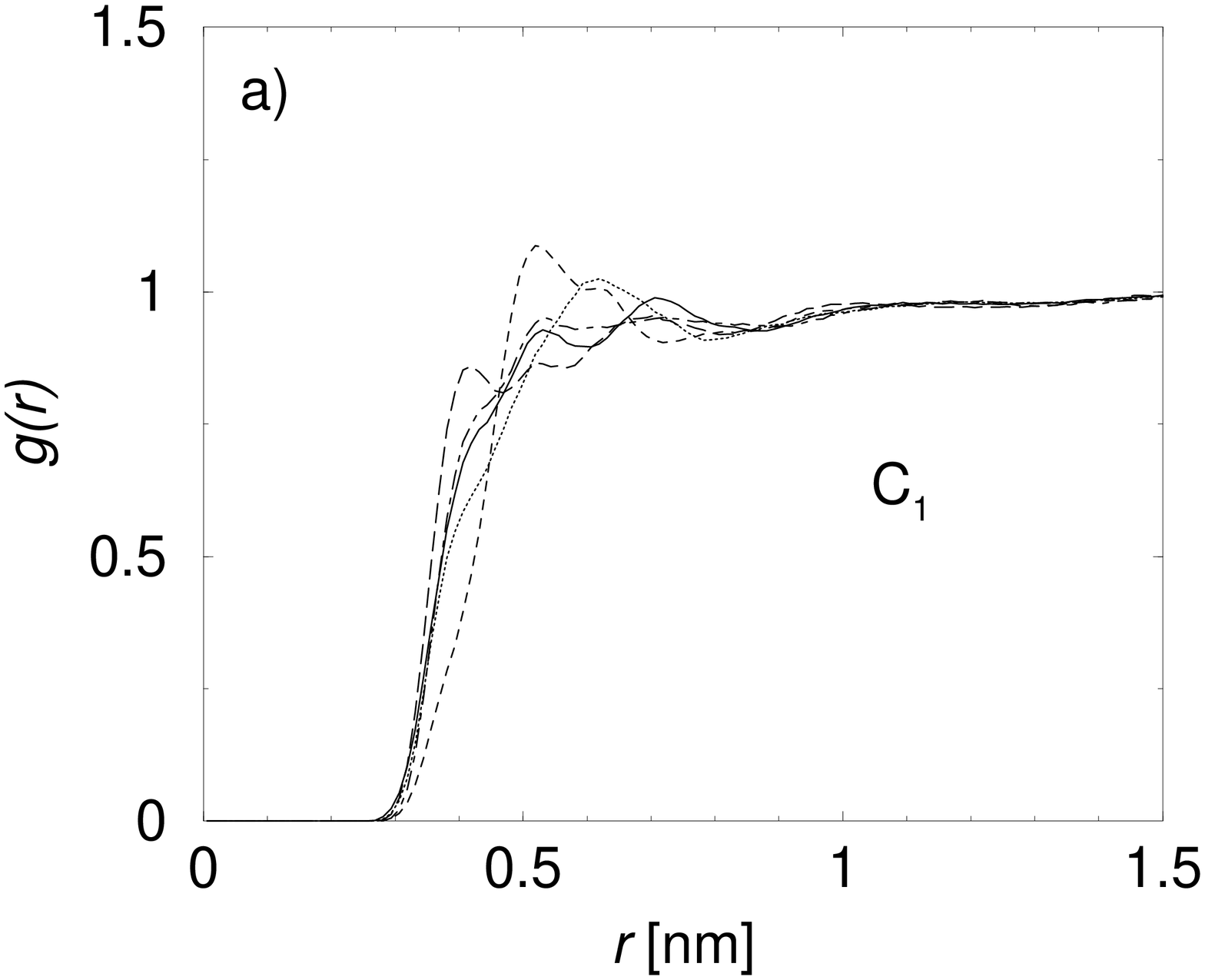}
  \includegraphics[angle=-90,width=0.5\linewidth]{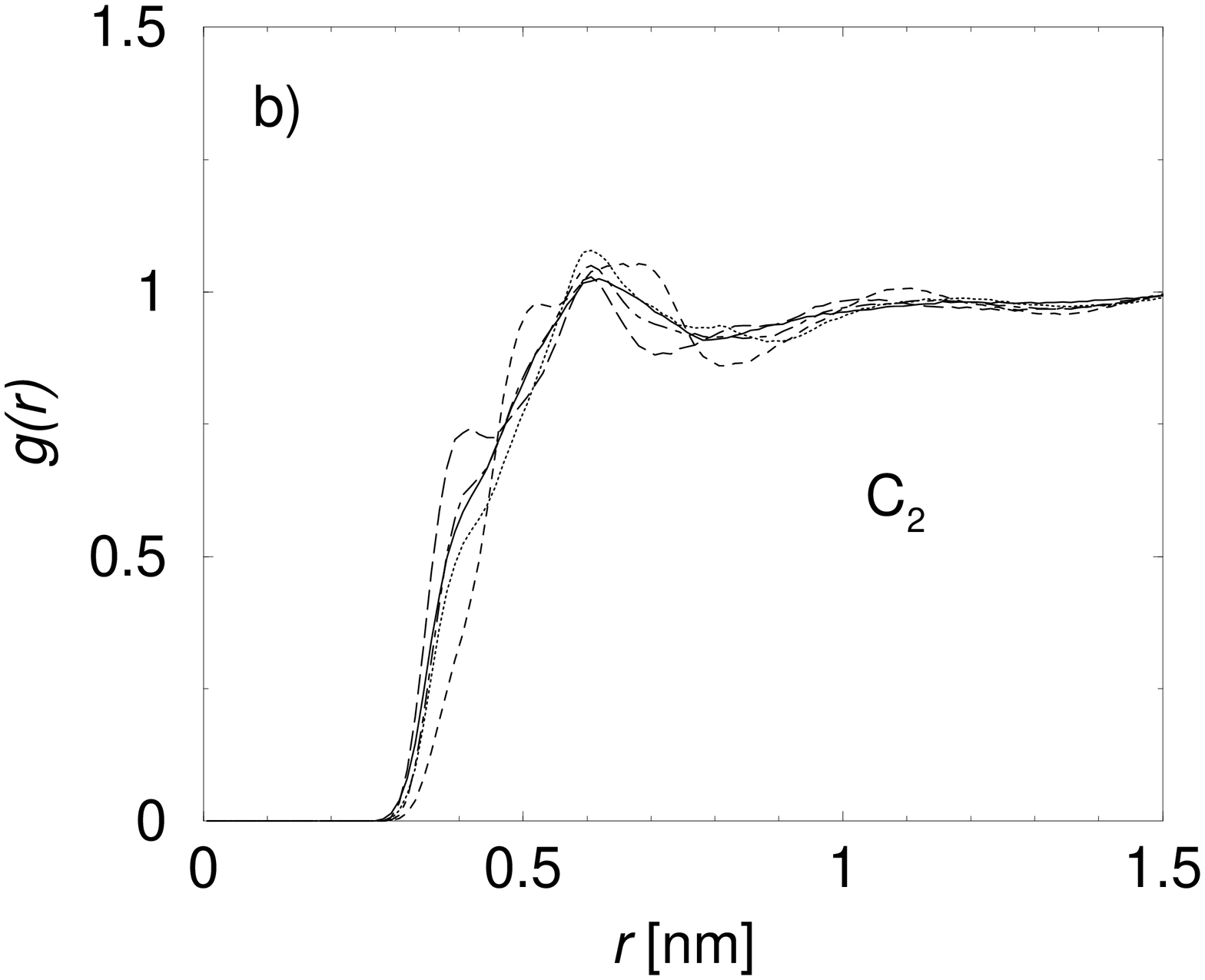}
  \includegraphics[angle=-90,width=0.5\linewidth]{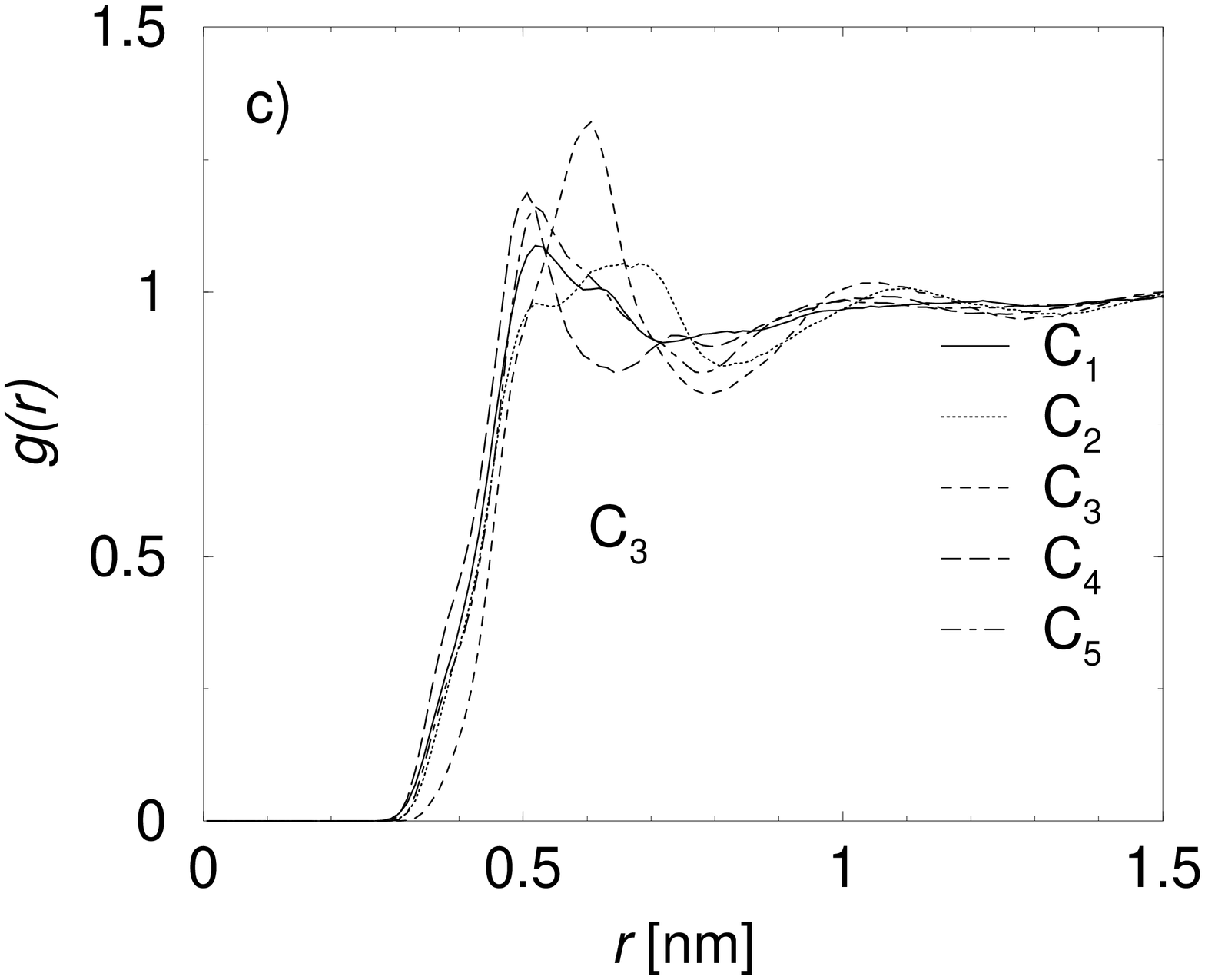}
  \includegraphics[angle=-90,width=0.5\linewidth]{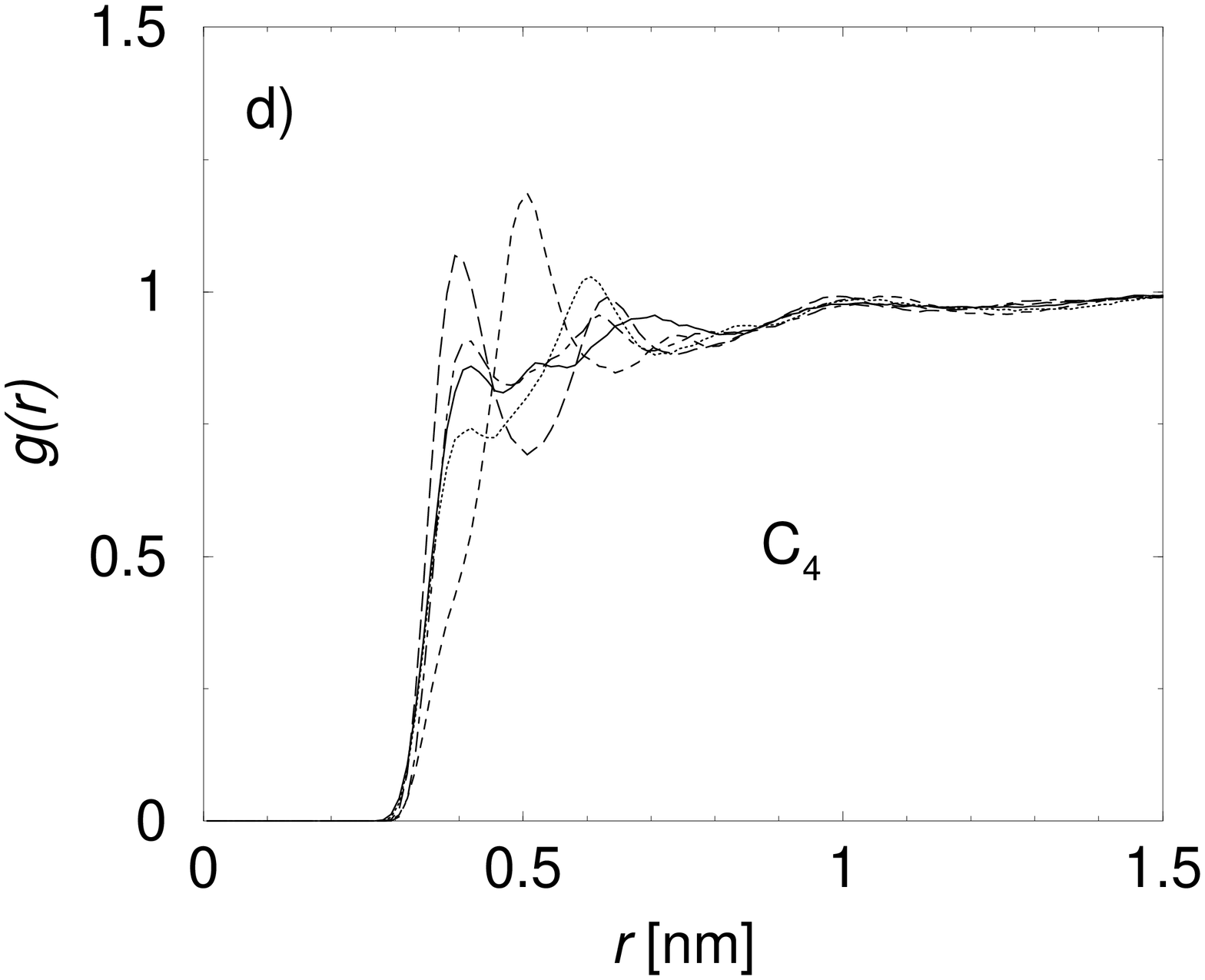}
  \begin{minipage}{0.5\linewidth}
    \includegraphics[angle=-90,width=\linewidth]{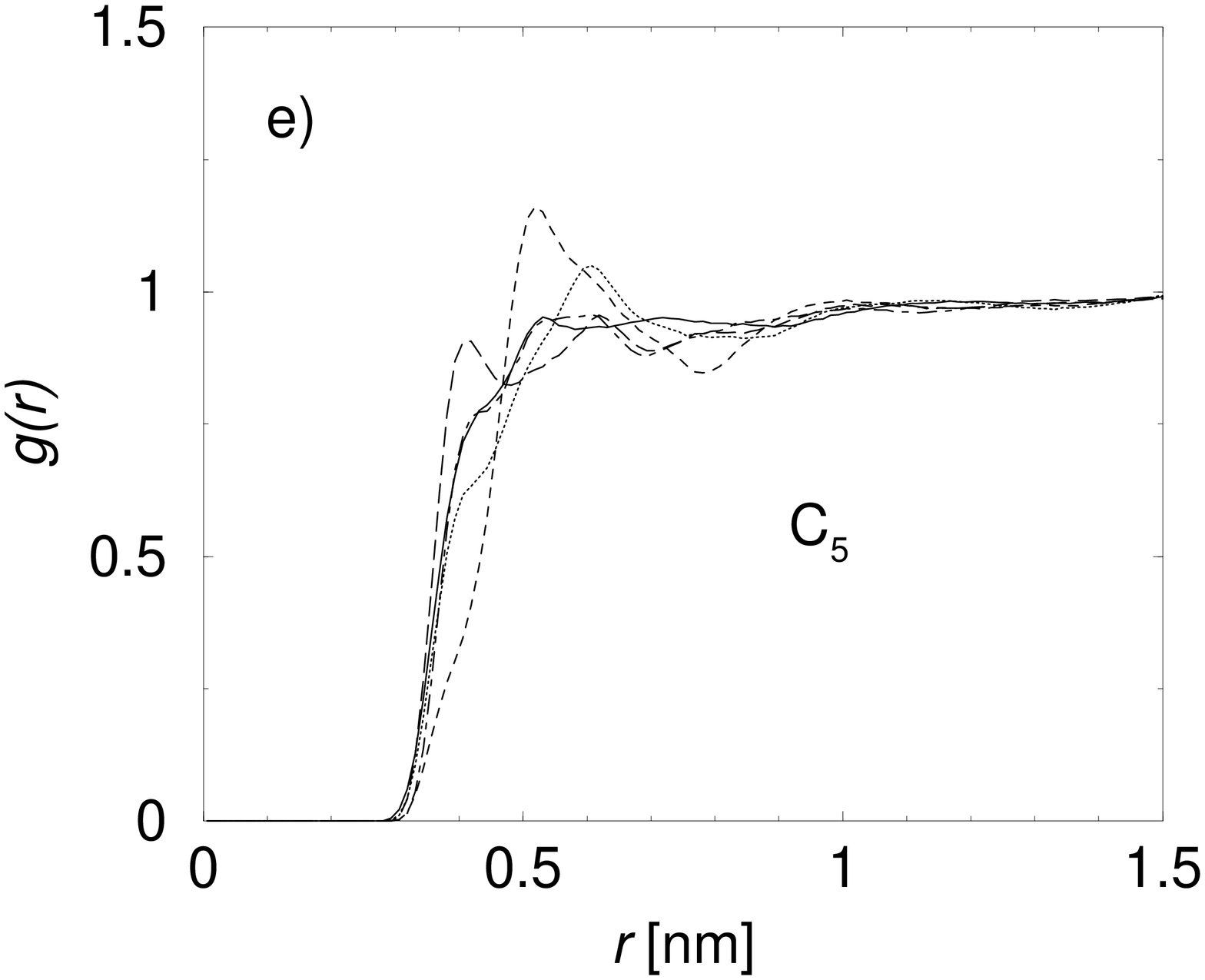}
  \end{minipage}
  \begin{minipage}{0.5\linewidth}
    \[
    \begin{array}{lrrrrr}
      & \opC_1 &  \opC_2 & \opC_3 & \opC_4 & \opC_5\\
      \opC_1 &  0.41  &  0.36   & 0.28   & 0.49   & 0.42  \\
      \opC_2 &  0.36  &  0.31   & 0.25   & 0.43   & 0.36  \\
      \opC_3 &  0.28  &  0.25   & 0.16   & 0.33   & 0.25  \\
      \opC_4 &  0.49  &  0.43   & 0.33   & 0.58   & 0.49  \\
      \opC_5 &  0.42  &  0.36   & 0.25   & 0.49   & 0.40  \\
    \end{array}
    \]
  \end{minipage}
  \caption{Partial inter-chain radial distribution functions (system~2,
    T=300~K). a) C$_1$ b) C$_2$ c) C$_3$ d) C$_4$ e) C$_5$. In all figures the
    definition of line styles in figure~c) applies.  Figure f (inset table):
    Integrated number of neighbors in the innermost shell
    $r<0.45\text{nm}$, only foreign chains.}
  \label{fig:partialrdf}
\end{figure}

Carbon$-$carbon radial distribution functions (RDF) resolved according to
carbon type allow the study of preferential arrangements between different
chains. Partial pair distribution functions between the five different carbons
present in polyisoprene were recorded (Figure~\ref{fig:partialrdf}).
The methyl carbon (C$_4$) is the most exposed and can, therefore, approach
closest to the others ($\approx0.4$nm). At this distance there is also
a shoulder in the C$_1$, C$_2$ and C$_5$-RDF indicating direct contact. C$_3$
is the most ``shielded'' carbon with a slight shoulder at direct contact to 
C$_1$. It is linked to C$_2$, C$_4$ and
C$_5$, thus, it is often found as second contact ($\approx0.55$nm). 
The two methylene carbons C$_1$ and C$_5$ are coordinated very
similarly. C$_2$ is easily accessible, as it
has only one hydrogen but not very exposed to other chains, since
it is part of the double bond in the backbone leading to weak structure.

Integration of the pair distribution function yields the number of neighbors
of an atom in a shell of radius $r$. By relating the local number density
$\rho_{\text{local}}=\frac{n_{\text{C}}}{4/3\pi r^3},\,n_{\text{C}}$ being the
number of carbons, to the overall concentration $\rho(\infty)=8.102$nm$^{-3}$,
local enrichment ($x:=\frac{\rho_{\text{local}}}{\rho(\infty)}>1)$ and
depletion ($x<1)$ can be resolved.
Overall integrated values in Figure~\ref{fig:partialrdf}f are smaller than 
unity due to the correlation hole, as only foreign chains are taken into 
account. In the innermost shell ($r<0.45$nm, cf. Figure~\ref{fig:partialrdf}f)
of all carbons methyl groups (C$_4$) are enriched, sp$^2$ carbons (C$_2$ and
esp. C$_3$) are depleted, whereas the methylenes (C$_1$ and C$_5$) are close
in concentration. In the second shell ($0.45<r<0.65$~nm), this is partly
reversed, as C$_3$ is enriched and C$_4$ is weakly depleted. At distances
larger than 0.65~nm all species are found at bulk concentration. In brief, one
can say that the methylene carbons occur at constant density almost
everywhere. The methyl and the double bonded carbons (esp. C$_3$) show much
more structure. Monomers of different chains thus approach each other
typically with their side groups as closest contact. Orientational influences
from the double-bond keeping the monomer planar play a role, too (see
below). The whole overall structure in the RDFs extends about two monomer
sizes ($r<1$~nm), whereas the concentrations of different carbons level out
already at 0.65~nm.

The local structure is not fully described by the (spherically averaged)
radial distribution functions. Mutual orientation is  
\begin{figure}
  \includegraphics[angle=-90,width=0.5\linewidth]{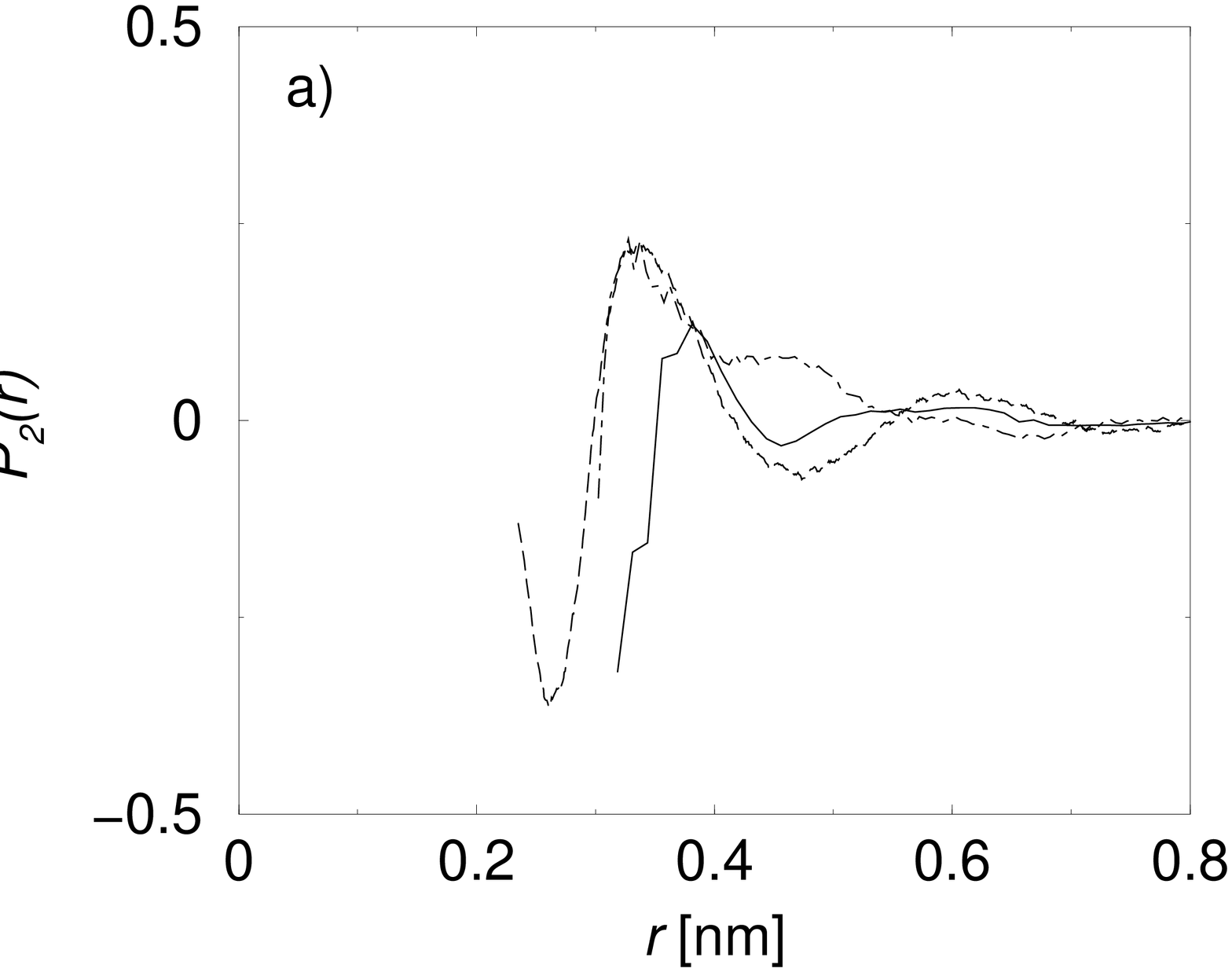}
  \includegraphics[angle=-90,width=0.5\linewidth]{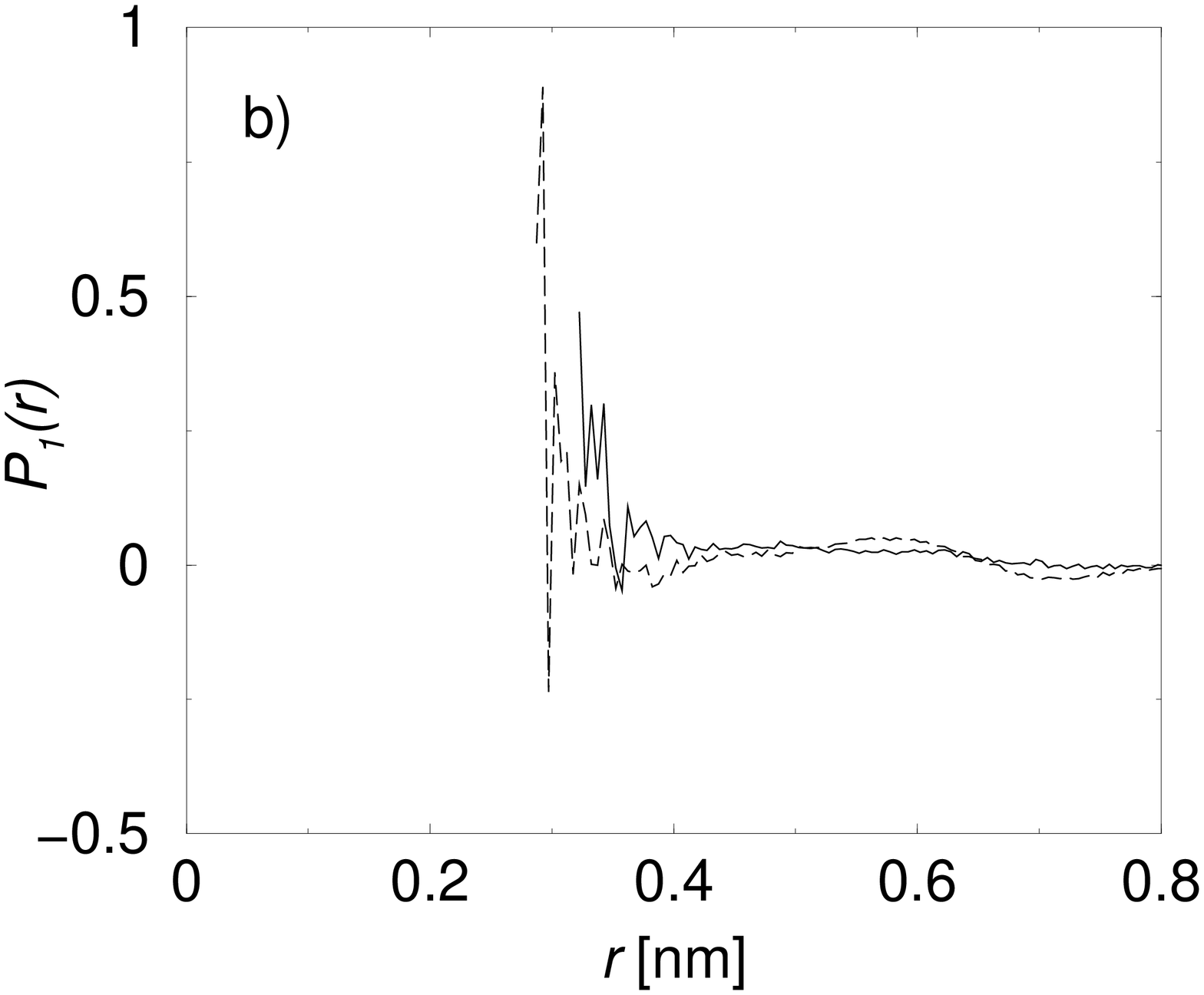}
  \includegraphics[angle=-90,width=0.5\linewidth]{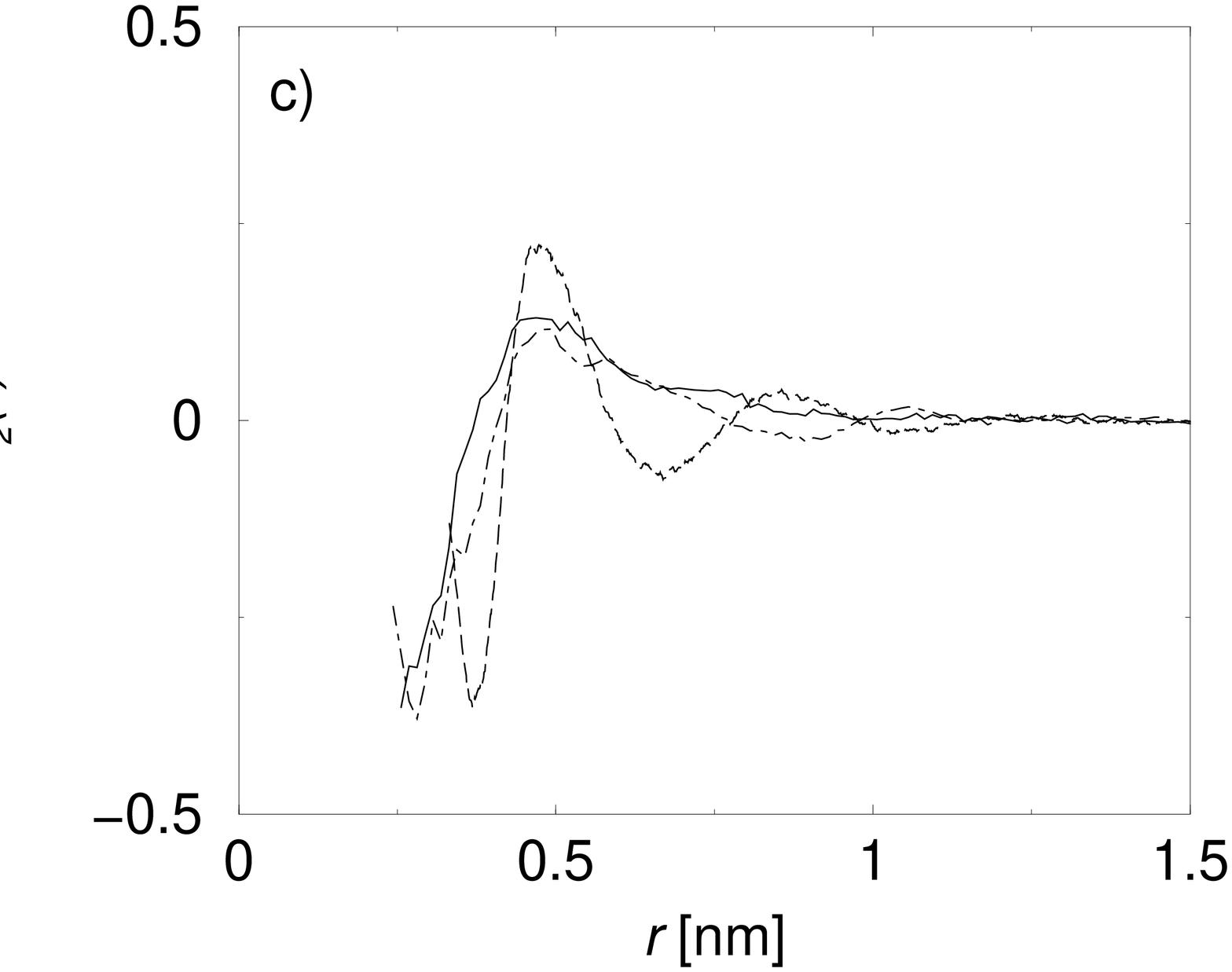}
  \caption{Static {\it inter} chain orientation correlations {\it OCF}
    (T=300~K, system~1).  a) Double-bonds and vectors connecting C$_5$ to
    C$_1$ of the next monomer between neighboring chains. Dot-dashed:
    atomistic double bond {\it OCF}; solid: C$_5-$C$_1$ vectors; dashed: {\it
    OCF} of a simple model with persistence length $l_p=1.5$ monomer
    units\cite{faller00b} scaled for coincidence at the first maximum to the
    solid one.  b) Same as figure~a) but $P_1$ in order to show the
    directionality of the correlations.  c) Inter-monomer vectors, solid line:
    C$_1-$C$_1$, dot-dashed line: C$_2-$C$_2$, dashed line: simple model for
    comparison. Note the different ordinate scale.}
  \label{fig:atoodf}
\end{figure}
measured by the orientation correlation function 
{\it OCF} (Figure~\ref{fig:atoodf})
\begin{equation}
  OCF(r)=P_{2}(r)=\frac{1}{2}\Big\langle
  3(\vec{u}_i\cdot\vec{u}_j)^2-1\Big\rangle\;. 
\end{equation}
The distance $r$ is measured between the centers of mass of the respective
pairs.  Again there are several
possible choices for the tangent vectors $\vec{u}_i$.
Orientation correlations of the vector connecting the methylenes (C$_1-$C$_5$)
extend over several inter-atomic distances (Figure~\ref{fig:atoodf}a). They
resemble the packing of model chains consisting of simple
beads~\cite{faller99b}.  The very few segments that approach closely
(cf. Figure~\ref{fig:rdf} and~\ref{fig:partialrdf}) show a perpendicular
orientation. A parallel ordering peak is encountered at about 0.4~nm. The
first Legendre polynomial ($P_1(r):=\langle\vec{u}_i\cdot\vec{u}_j\rangle$,
Figure~\ref{fig:atoodf}b) which carries direction information shows that these
inter-chain contacts have a small preference of
neighboring chains running in the same direction. At larger distances the
explicit atomistic structure is no longer important and the structures become
much broader. But there are still structural effects originating from the
packing visible up to about 0.7~nm, about two chain diameters.

The order of the double bonds between the chains exhibits more atomistic
details. As the double bond lead to a planar configuration of the environment,
there is parallel orientation of the neighbors especially at short
distances ($r\lesssim0.4$~nm) which is visible in the first Legendre
polynomial. The bond vectors C$_5-$C$_1$ (between subsequent monomers), which 
are about the same length as the double bonds, show even less structure than 
the simple model.

The inter-monomer vectors (Figure~\ref{fig:atoodf}c), in contrast, display even
more generic features which can also be seen in a simple
model~\cite{faller99b,faller00b,mplathe00} (compare
Section~\ref{sec:simple}). As they describe larger segments compared to the
vectors discussed above, the features are less accentuated.  On close contact
($r\lesssim0.4\text{nm}$) an almost perfect perpendicular order is found. At
intermediate distances ($0.4\text{nm}<r<0.5\text{nm}$) a preferred parallel
alignment, and for the case of the C$_2-$C$_2$ vector a second perpendicular
region appears ($r\approx0.8\text{nm}$). The differences between the two
different inter-monomer (C$_1-$C$_1$ and C$_2-$C$_2$) vectors are small. So
for orientations on length scales as small as monomer sizes, simple models
already provide a useful description if the persistence lengths are the same. 

This shows that the generic packing effects are important for the
understanding of the structure of atomistic models. However, the fine
structure at short distances, as found in the first Legendre polynomial,
cannot be explained by generic arguments as here the detailed chemistry of the
polymer is important.
\subsection{Structure of the melt}
Radial distribution functions are quite illustrative in characterizing the
structure. However, the experimental observable in scattering experiments is
the structure function. The static melt structure factor
\begin{figure}
  \begin{center}
    \includegraphics[angle=-90,width=0.4\linewidth]{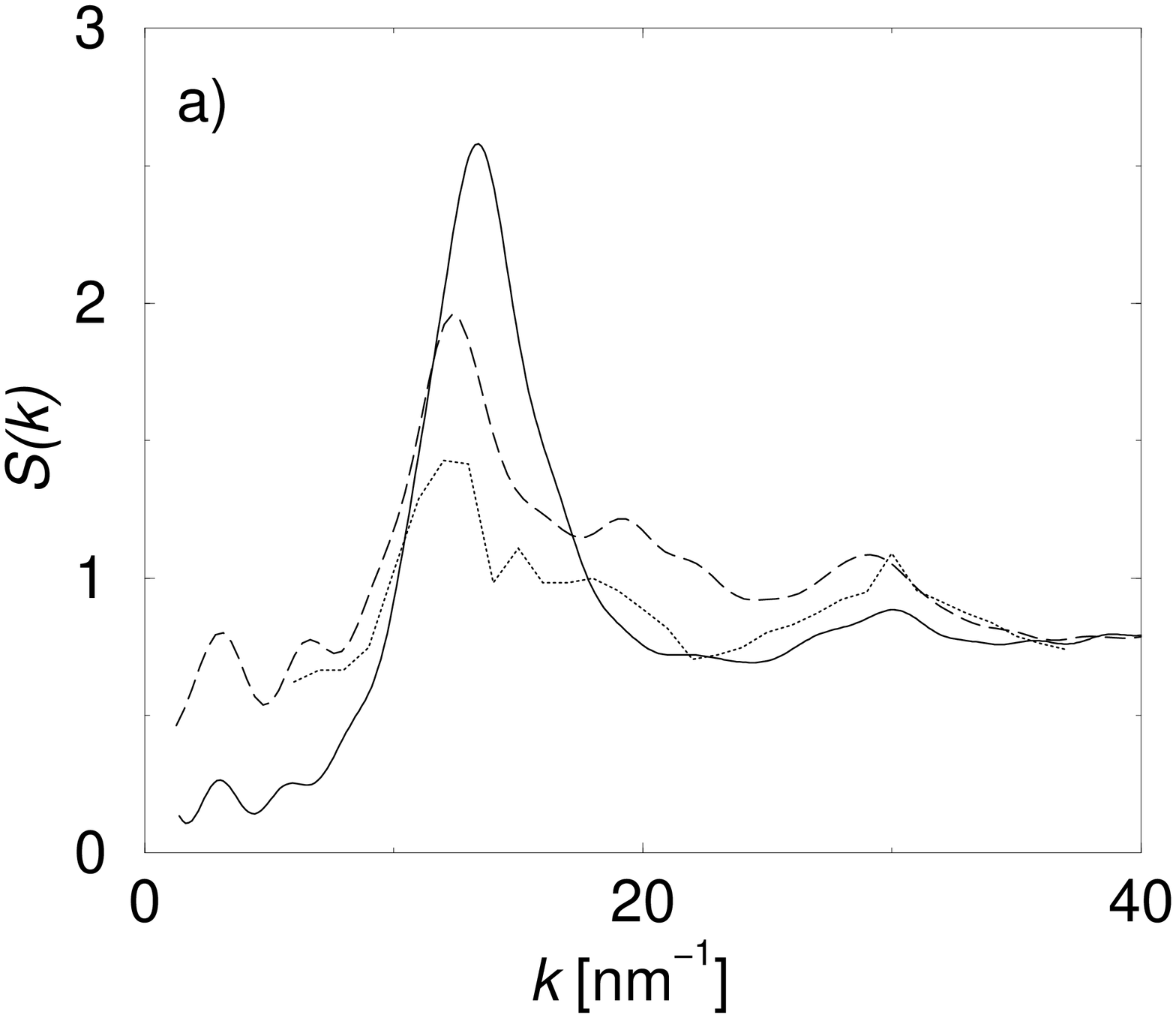}
    \includegraphics[angle=-90,width=0.4\linewidth]{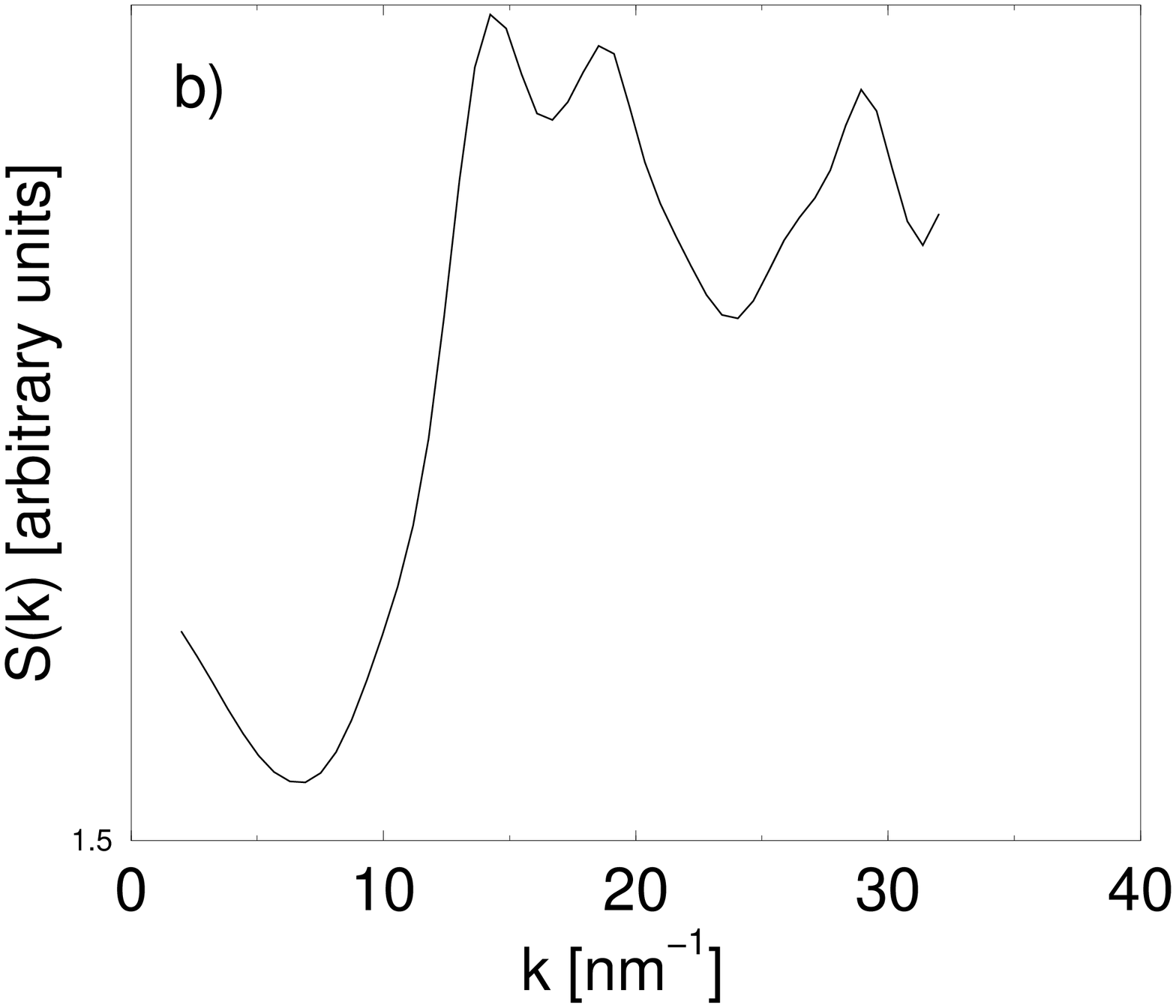}
    \caption{a) Static structure factor of {\it trans} polyisoprene from this
      work (solid line: T=300~K, dashed line T=413~K) with the scattering
      lengths of all atoms taken to be the same in comparison to cis-PI
      (dotted line, T=413~K, data from ref.~\citen{moe99}). b) Experimental
      structure factor obtained by neutron scattering at an unknown
      temperature at molecular weight of 17.200 (data from
      ref.~\citen{zorn92}).}
    \label{fig:strfct}
  \end{center}
\end{figure}
\begin{equation}
  S_{\text{melt}}=  
  \frac{1}{N}\Bigg\langle\left|\sum_{m=1}^{N_{\text{C}}}\sum_{j=1}^{N} 
    \exp(ikr_j^m)\right|^2\Bigg\rangle
\end{equation}
is shown in Figure~\ref{fig:strfct} in comparison to simulations of {\it
cis}-polyisoprene~\cite{moe99} and experiments on a mixture dominated by the
{\it cis}-conformer~\cite{zorn92}. The melt structure factor shows a clear
peak at about 15~nm$^{-1}$. In addition, there is some smaller structure,
especially higher order peaks. The lower limit of resolution is given by the
size of the box corresponding to a minimum
$|\vec{k}|$-vector of about 0.4$\pi$~nm$^{-1}$. At higher temperature the
overall structure flattens out with less pronounced peaks, as expected.

Moe and Ediger performed simulations on pure {\it cis}-polyisoprene at 363~K
and 413~K with one chain of 100 monomers~\cite{moe96a,moe99}. This, obviously,
reduces the influence of end effects. The densities were much lower than the
experimental values (798 vs. 869 and 775 vs. 836~kg/m$^3$, respectively). The
structure functions of both simulations are, however, comparable at
413~K. Neither simulation compares really satisfactorily to the experimental
structure factor. The double-maximum structure of the first peak is not
reproduced, the lower maximum ($\approx$13~nm$^{-1}$) being enhanced, the
higher ($\approx18$~nm$^{-1}$) being reduced to a shoulder. The positions of
the peaks, however, are in reasonable agreement. As the experimental 
temperature is
not given, one cannot say whether at least part of the discrepancy is a low
temperature effect, indicating the formation of a glass, whereas both
simulations are performed deep in the melt. Another reason for the
disagreement could be the fact that the short chains presented are still
oligomers. The alternative would be one long chain, as in refs.~\citen{moe96a}
and~\citen{moe99}. However, the periodic boundary replication of a single
chain this imposes a strong artificial periodicity to a amorphous melt, which 
makes it difficult to look at {\it inter}-chain effects.
\section{Dynamical properties}
\subsection{End-to-end vector}
\begin{figure}
  \includegraphics[angle=-90,width=0.5\linewidth]{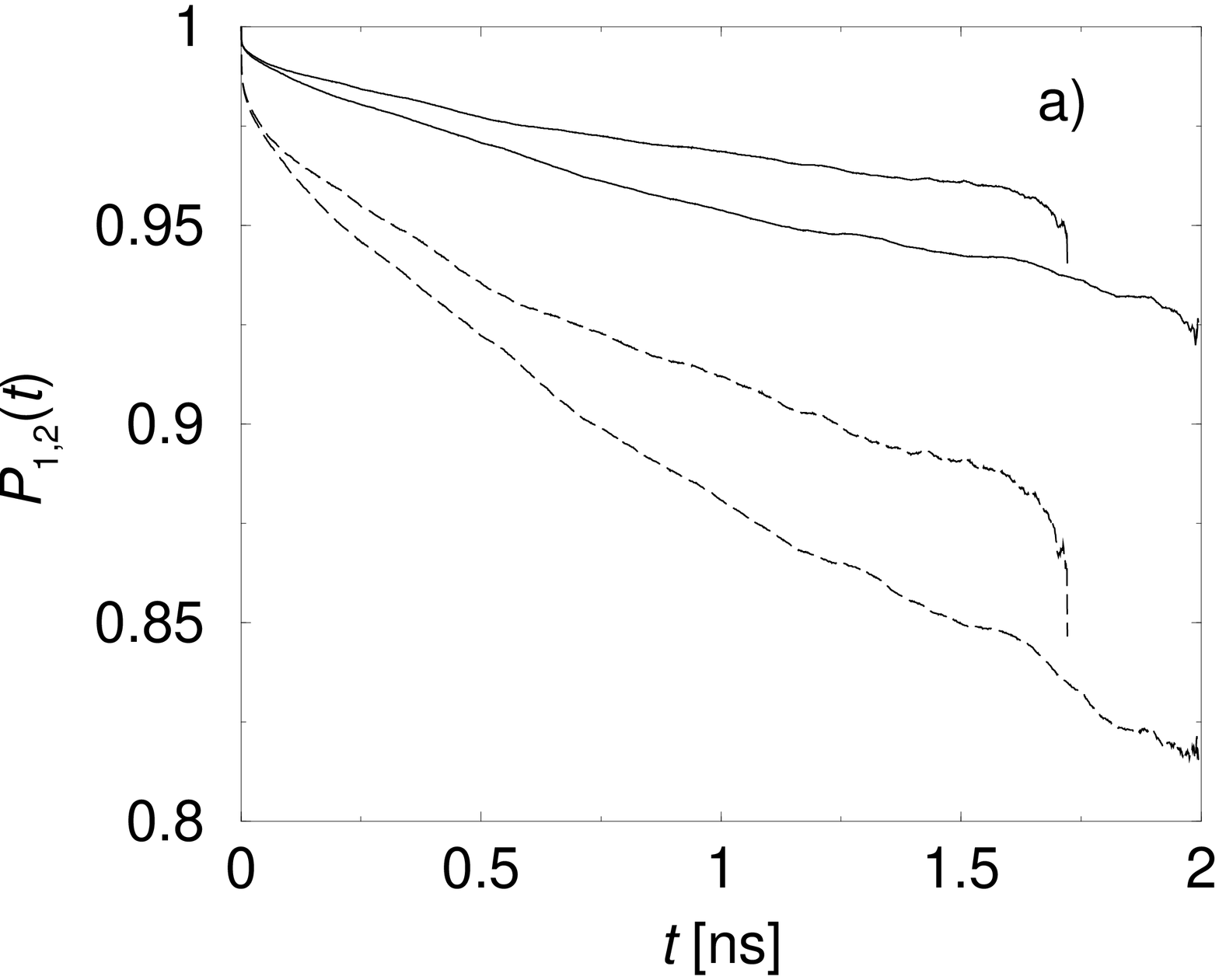}
  \includegraphics[angle=-90,width=0.5\linewidth]{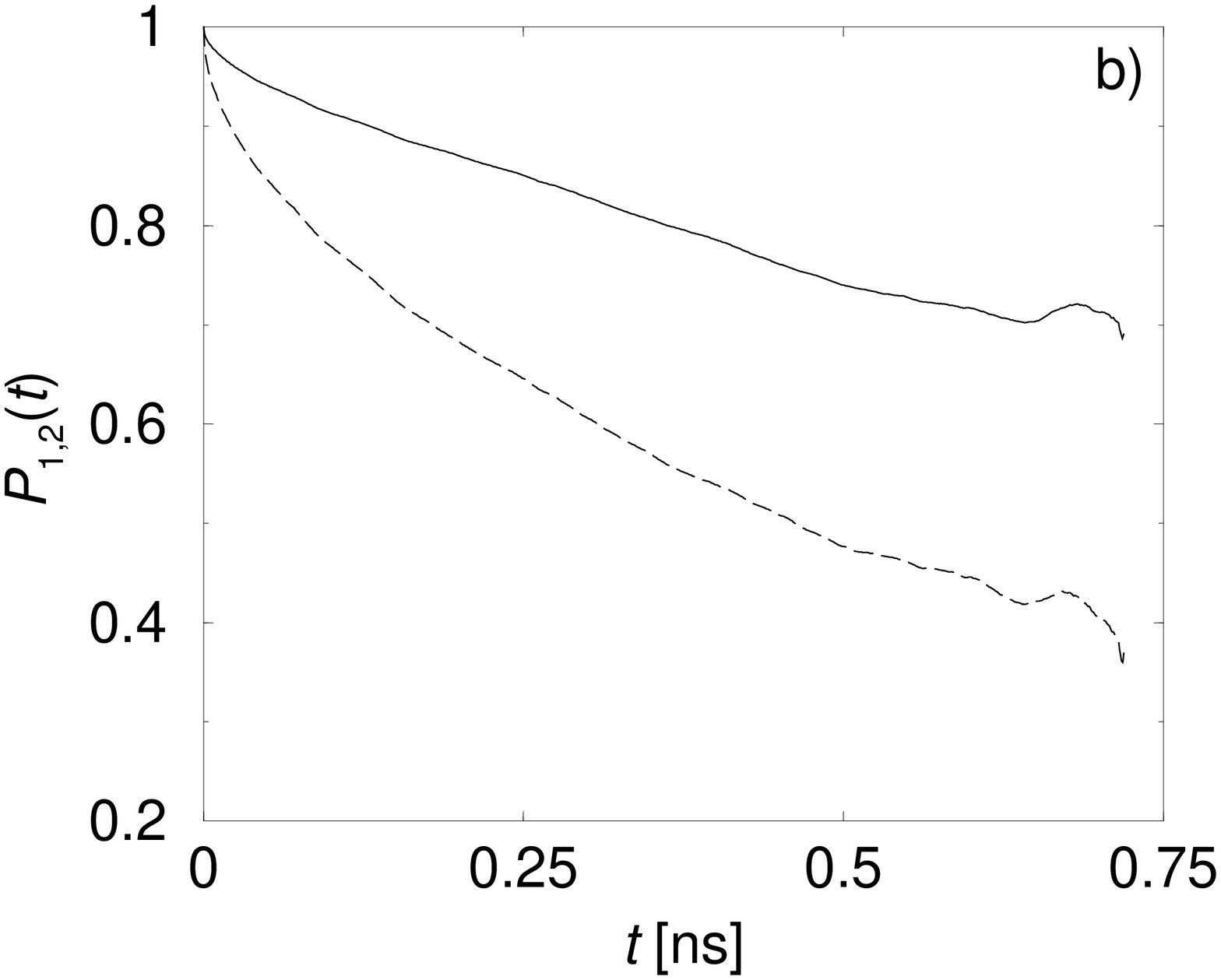}
  \caption[Reorientation of atomistic end-to-end vector]
  {Relaxation of the end-to-end vector. Solid curves are first Legendre
    polynomials, dashed curves second Legendre polynomials. 

    a) T=300~K: Of both pairs, the upper curve is the correlation
    function for system~3 and the lower one for system~2. 

    b) For T=413~K (system~3) the relaxation times are much shorter. 
    } 
  \label{fig:end-end}
\end{figure}

Figure~\ref{fig:end-end} shows for the two polydisperse systems~2 and~3 the
reorientation correlation functions (first and second Legendre polynomial)
\begin{eqnarray}
  P_1(t)&=&\Big\langle(\vec{u}(t)\cdot\vec{u}(0))\Big\rangle\\
  C_{\text{reor}}(t):=P_2(t)&=&\Big\langle0.5\big[
  3(\vec{u}(t)\cdot\vec{u}(0))^2-1\big]\Big\rangle
\end{eqnarray}
of the end-to-end vector of the chains of length $N=10$, which is defined as
the vector connecting the two terminal carbons C$_1^{\text{mono 1}}$ and
C$_5^{\text{mono }n}$.  The relaxation time is clearly longer than the time
accessible in the simulations. Even system~2, which was simulated for more
than 2~ns, did not relax appreciably on this time scale; local vectors relax,
of course, much faster. At the higher temperature of 413~K the relaxation time
decreases drastically. Still, one has to be cautious discussing length scales
of more than a monomer.
This figure shows some scatter between the two systems, which may
be taken as a rough estimate of the error of the simulations.
\subsection{C$-$H bond reorientation}
Reorientation in polyisoprene melts with  92\% {\it cis}-conformer was measured
by the group of Laupr\^etre and Monnerie~\cite{batie89,laupretre93}. Another 
investigation with a higher {\it trans}
content of 22\%~\cite{denault89} focused also on the {\it
cis}-conformer. Experimentally, the direct observable is the $T_1$ time.
\begin{equation}
  \frac{1}{T_{1}}=\frac{\hbar^{2}\gamma_{\opC}^{2}
    \gamma_{\opH}^{2}}{10r^{2}_{\text{C}-\text{H}}} 
  \Big[J(\omega_{\opH}-\omega_{\opC})+3J(\omega_{\opC})+
  6J(\omega_{\opH}+\omega_{\opC})\Big].
\end{equation}
The $\gamma_{i}$ are the gyro-magnetic ratios of the respective nuclei and
$\omega_{\opC}$ and $\omega_{\opH}$ are the Larmor frequencies, $r$ is the
distance between the nuclei. The function $J(\omega)$ is the spectral density,
i.e. the Fourier transform of $C_{\text{reor}}$ of the respective C$-$H vector
\begin{equation}
  J(\omega)=\frac{1}{2}\int_{-\infty}^{\infty}C_{\text{reor}}(t)e^{i\omega t}
  \text{d}t\;.
\end{equation}

In atomistic simulations, $T_{1}$ has been determined for different
polymers\cite{moe96a,paul97,antoniadis98}. Moe and Ediger use the limit of
{\it extreme narrowing} ($\omega\tau_{\text{reor}}\ll1$) to analyze their {\it
cis}-polyisoprene data at $\text{T}=413$~K, as is done in most other atomistic
simulations\cite{moe96a}. This has the advantage that $T_{1}$ becomes
independent of the Larmor frequencies\cite{kalinowski84}, which
cannot be measured in simulations without extrapolation. For long
chains in a simple model one sees that extrapolation starting at such high
values is very questionable.\cite{faller00b} The spectral density $J(\omega)$
is for very short reorientation times independent of $\omega$:
\begin{equation}
  J(\omega)=B_{\text{local}}^2\frac{2\tau_{\text{reor}}}
  {1+\omega^2\tau_{\text{reor}}^2}\;.
\end{equation}
However, the {\it extreme narrowing} regime is
not reached normally by the experiments, as high temperatures are needed to
yield correlation times that the results correspond to the limit.  

The reorientation time $\tau_{\text{reor}}$ is defined by the time integral
over the correlation function
\begin{equation}
  \tau_{\text{reor}}=\int_0^{\infty}C_{\text{reor}}\text{d}t\;,
\end{equation}
and in the extreme narrowing limit this is directly linked to the  $T_1$ time
for a C$-$H vector
\begin{equation}
  T_1^{-1}=10nK\tau_{\text{reor}}\;,
\end{equation}
where $K$ is a constant related to the bond length and $n$ is the number of
protons connected to the respective $^{13}$C. Note that a shorter $T_1$
corresponds to a longer reorientation time.

\begin{figure}
  \includegraphics[angle=-90,width=0.5\linewidth]{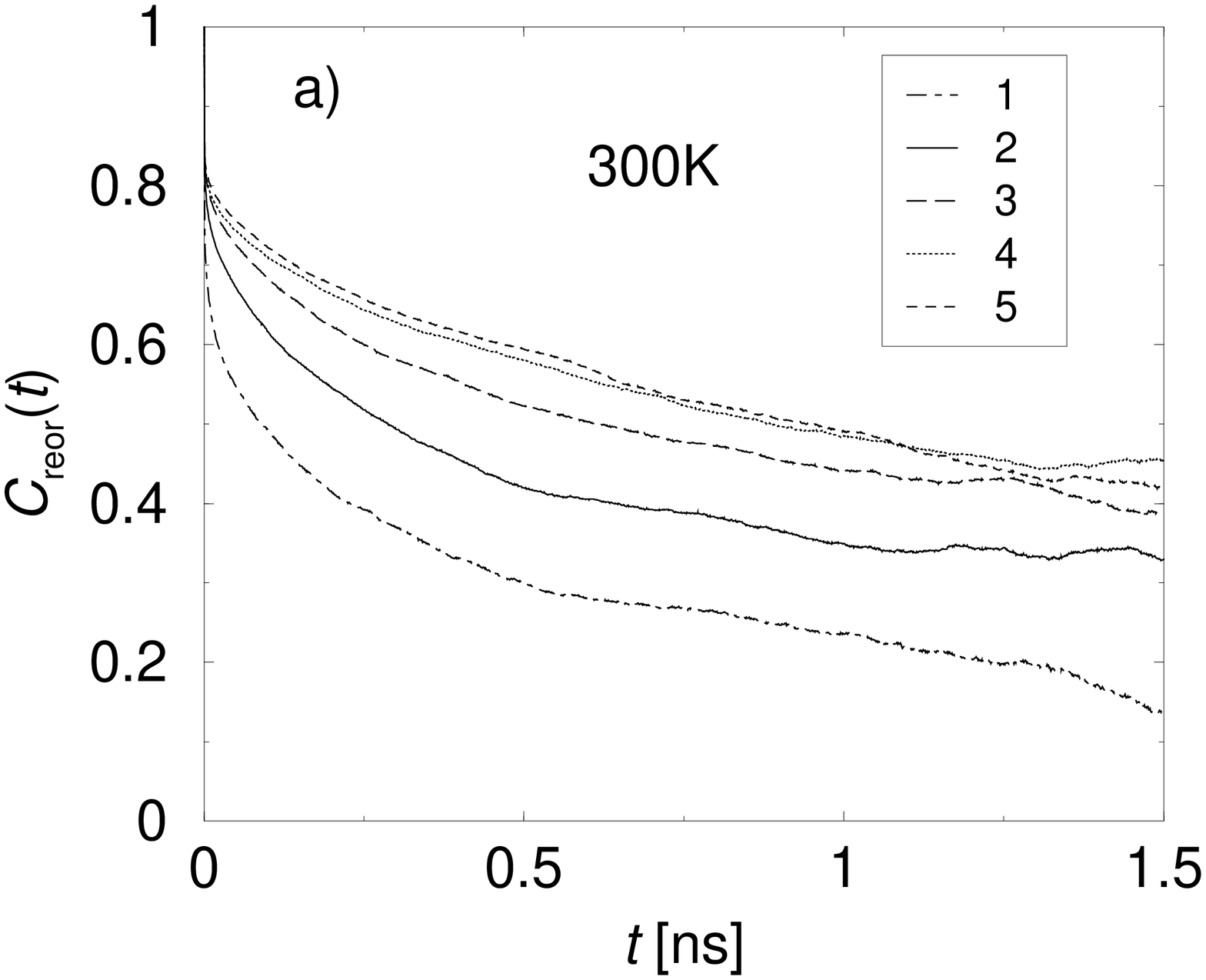}
  \includegraphics[angle=-90,width=0.5\linewidth]{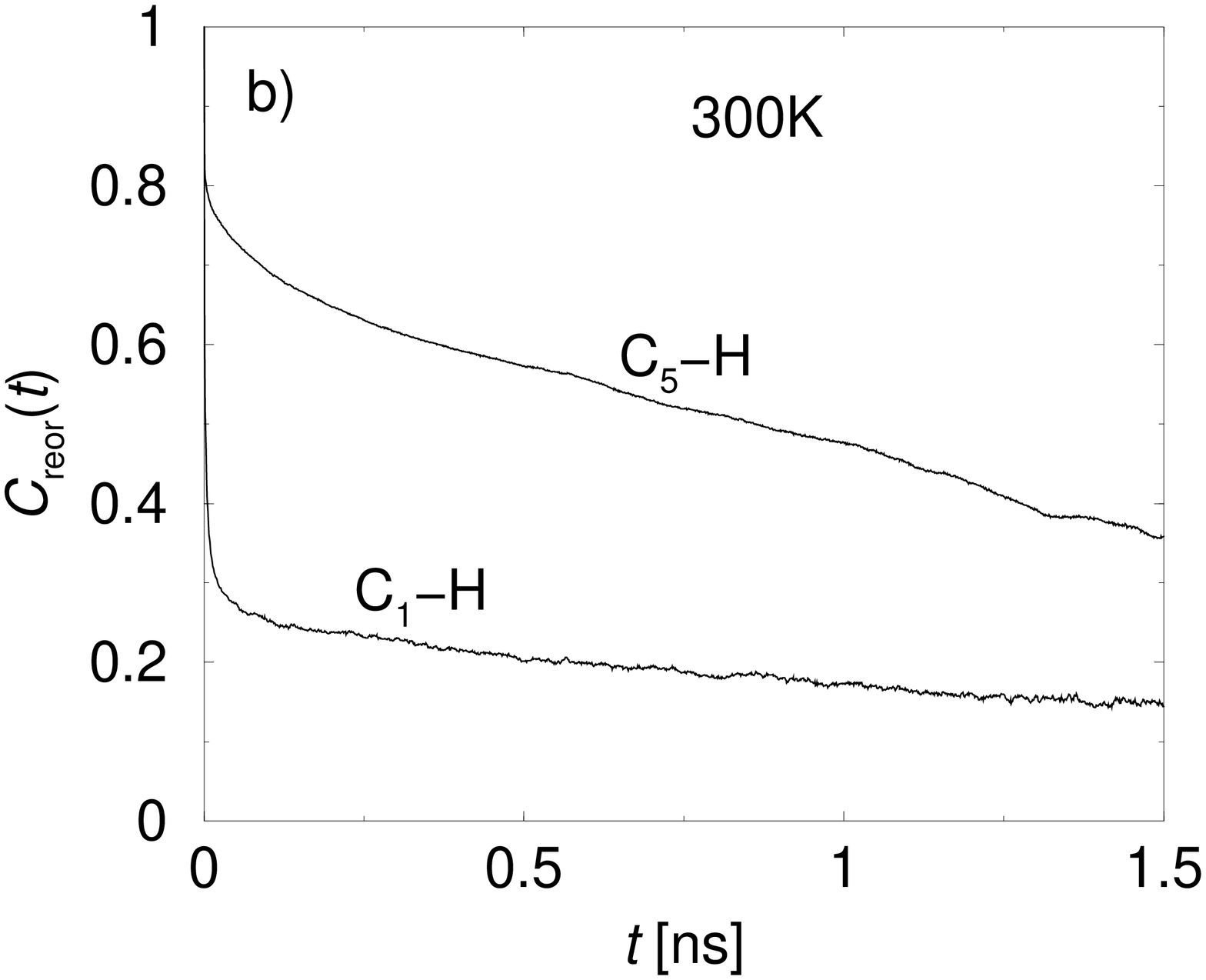}
  \includegraphics[angle=-90,width=0.5\linewidth]{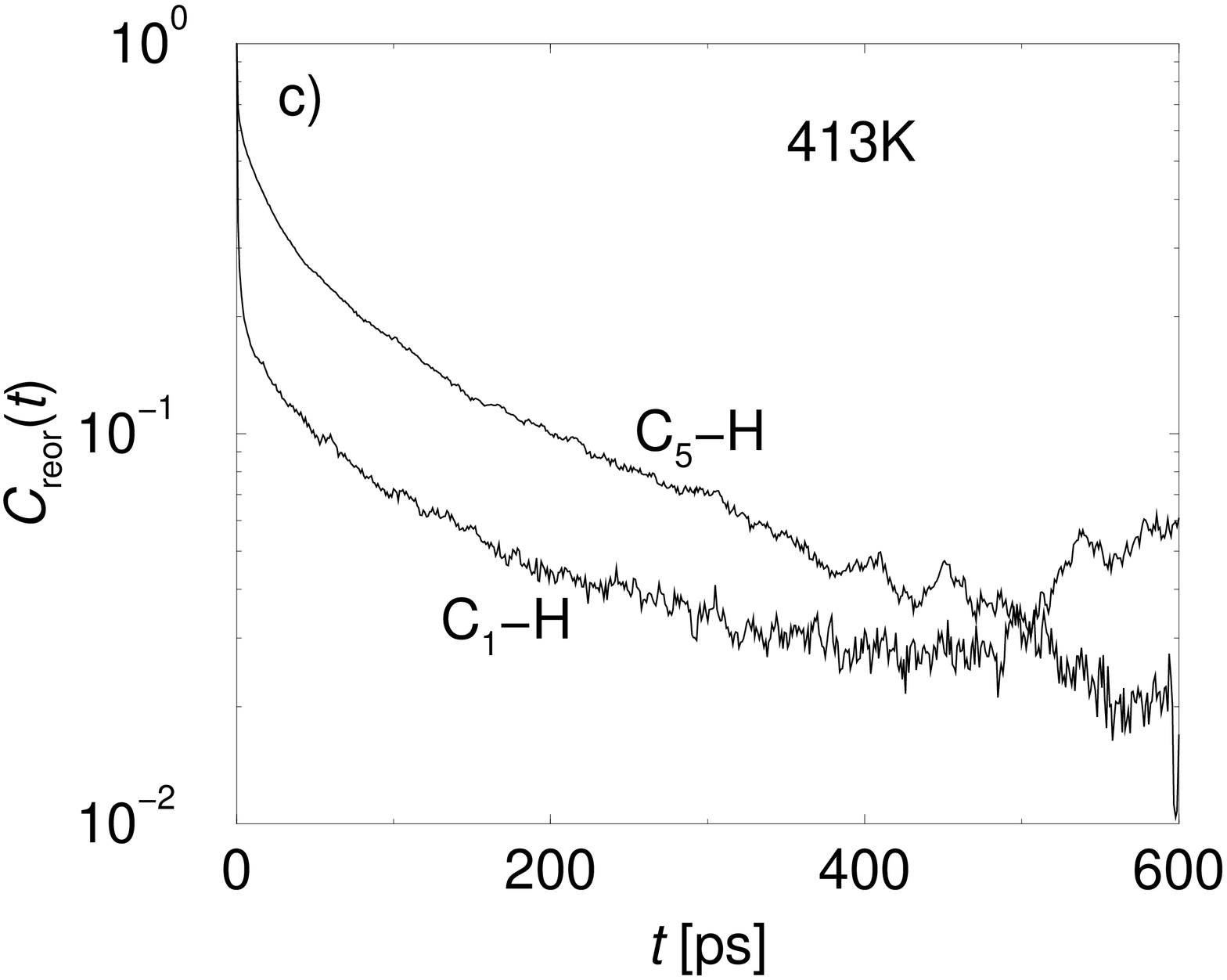}
  \caption{Reorientation of C$-$H vectors in inner monomers of atomistic
    polyisoprene chains (system~3, chains of length~10,) a) Vinyl C$_2-$H
    depending on monomer position (1: end monomer, 2: next to end monomer,
    etc.), b, c) Methylene groups of the central monomer at different
    temperatures.} 
  \label{fig:CHreor}
\end{figure}

The C$-$H vector reorientation is followed in the simulations. To minimize
chain-end effects only the inner-chain monomers are included in
Figure~\ref{fig:CHreor}~b and~c.
Comparing Figures~\ref{fig:dblbndreor}~a and \ref{fig:CHreor}~a, one sees that
the hydrogen connected to the backbone at carbon C$_2$ is strongly tied to the
backbone. Thus, even such local quantities as bond vectors can be used as 
observables to monitor the dynamics of intermediate-size chain segments. 

Figure~\ref{fig:CHreor}~b illustrates that different torsion potentials and the
side group have considerable influence on the reorientation of vectors. The
reorientations of the two methylene carbons C$_1$ and C$_5$ differ. As C$_5$ 
is vicinal to the methyl
group, the steric hindrance for the hydrogens tied to it is more
pronounced. Thus, the initial stage of the reorientation,  present for
the C$_1-$H vector, is considerably smaller for C$_5-$H. This
is still visible at 413~K (Figure~\ref{fig:CHreor}~c), although all relaxations
are faster. The above-mentioned experiments~\cite{batie89,denault89} on {\it
cis}-polyisoprene melts show the same tendenies.

Similar to our reorientation correlation function
(cf. Figure~\ref{fig:CHreor}) the experimental data~\cite{batie89} provide
evidence for a two stage process, the first part being simply exponential
followed by a non-exponential long-term stage, described by model correlation 
function
\begin{equation}
  C_{\text{reor}}(t) =ae^{-t\tau_0} + (1-a)e^{-t/\tau_2}e^{-t/\tau_1}
  I_0(t/\tau_1)\;,
  \label{eq:modelreor1}
\end{equation}
with $\tau_0$ the local libration time, $\tau_1$ the time of conformation
jumps, and $\tau_2$ connected to damping; $I_0$ is a Bessel function. The
separation of time scales for polyisoprene is $\tau_1/\tau_0\ge150$ for the
two faster characteristic times~\cite{laupretre93}; the two slow processes
($\tau_1,\,\tau_2$) differ by a factor of 40. A separation of motions was used
by Lipari and Szabo as well to analyze NMR data of
polymers~\cite{lipari82a,lipari82b}. The correlation function they used has
the simpler double exponential shape
\begin{equation}
  C_{\text{reor}}(t)={\cal S}^2e^{-t/\tau_1}+(1-{\cal S}^2)e^{-t/\tau_2}\;,
  \label{eq:12}
\end{equation}
where the generalized order parameter ${\cal S}$ is related to the parameter
$a$ of Laupr\^etre {\it  et. al.}~\cite{laupretre93}. The
reorientation motion is described by one local and one global reorientation if 
there is no reptation.

Actual numbers for the three times of the model (Eq.~\ref{eq:modelreor1}) are
not provided in refs.~\citen{batie89,laupretre93}, but only values for
$a$. Although the experimental polymer is mainly {\it cis} the values for $a$
for the different vectors are comparable between our simulations and
experiment (Table~\ref{tab:avalues}). The simulation data in our case were
determined using a fit to Eq.~\ref{eq:12} with a purely exponential form of
the second term disregarding the Bessel function ($I_0=1$). The second column
in Table~\ref{tab:avalues} shows the values of $a$ estimated from the value of
$C_{\text{reor}}$ at 1~ps, which is the shortest time resolved in the
simulations ($a_{1\text{ps}}=1-C_{\text{reor}}(1\text{ps})$), coming closer to
experimental values. This assumes that the first process is too fast to be 
resolved here.  If the conformational jump time is
regarded as the time for torsion rearrangement (see below), and keeping in
mind that the two times differ at least by a factor of 150, the guess for
$a_{1\text{ps}}$ is probably more realistic.
\begin{table}
  \[
  \begin{array}{c*{4}{D{.}{.}{-1}}}
    \hline
    \text{vector} 
    & \mul{a_{\text{fit}}^{trans}} & \mul{a_{1\text{ps}}^{trans}} &  
    \mul{a_{\text{sim}}^{cis}}& \mul{a_{\text{exp}}^{cis}}\\ 
    \hline
    & \multicolumn{2}{c}{\text{sim: this work}} 
    & \mul{\text{sim: ref. \citen{moe99}}} & 
      \mul{\text{exp: ref. \citen{batie89}}} \\
    \hline
    \operatorname{C}_{1}-\operatorname{H} & 0.75 & 0.42 & 0.28  & 0.40\\
    \operatorname{C}_{2}-\operatorname{H} & 0.29 & 0.16 & 0.16  & 0.17\\
    \operatorname{C}_{5}-\operatorname{H} & 0.29 & 0.18 & 0.23  & 0.48\\
    \hline
  \end{array}
  \]
  \caption{Comparison of the experimental ({\it cis}-PI) and simulation ({\it
      cis} and {\it trans}-PI, system~2) data for the efficiency of the
    initial stage of the reorientation process. $a_{\text{fit}}$ originates
    from an exponential fit of the second stage extrapolated to $t=0$ and
    $a_{1\text{ps}}$ is the value of $1-C_{\text{reor}}$ at 1ps.  In the
    analysis of the simulations for {\it cis}-polyisoprene a stretched
    exponential second process was assumed. The experiments used a range of
    temperature between 283~K and 363~K (ref.~\citen{batie89}). The {\it
      trans} simulations were at 300~K (this work) and the {\it cis}
    simulations at 363~K (ref.~\citen{moe99}).}
  \label{tab:avalues}
\end{table}
Still, the simulations underestimate the difference between C$_1$ and C$_2$
and overestimate the one between C$_1$ and C$_5$. The data provided here 
correspond to a sample of pure {\it trans}-polyisoprene
oligomers which is presumably quite different from real {\it cis}-polyisoprene
with some added {\it trans}-conformer. Moreover, the discrepancy becomes
weaker for system~1 (below). Recent simulations on {\it cis}-polyisoprene at
higher temperature (T=363~K and T=413~K) were interpreted in terms of a two
stage correlation, too. There a separation between the exponential first
stage and a stretched exponential second process was deduced~\cite{moe99}. The
corresponding $a$-parameters are included in Table~\ref{tab:avalues}. Except
for the C$_5-$H vector they are comparable to our data. For
C$_5-$H they are even farther from the experimental value.

\begin{figure}
  \includegraphics[angle=-90,width=0.5\linewidth]{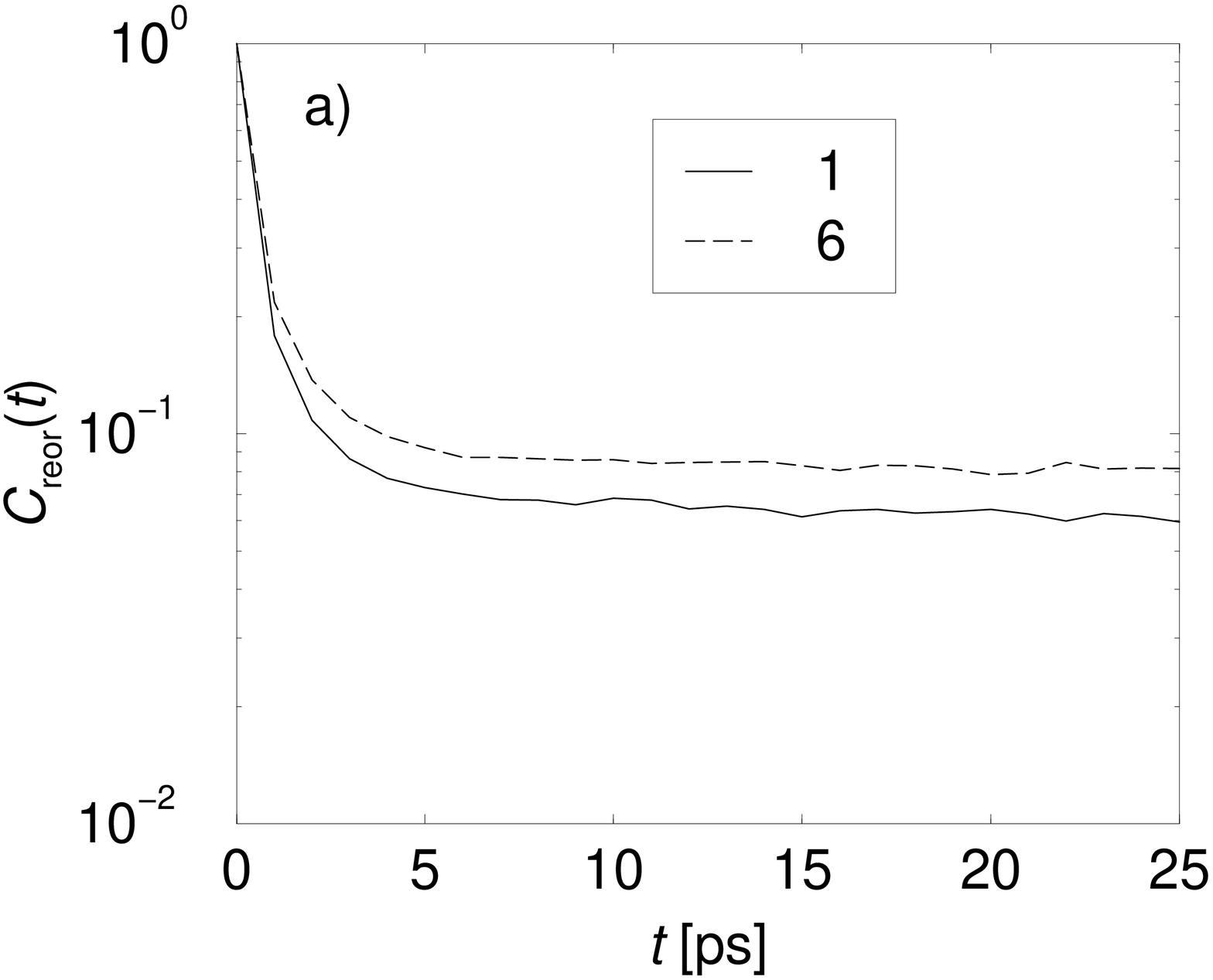}
  \includegraphics[angle=-90,width=0.5\linewidth]{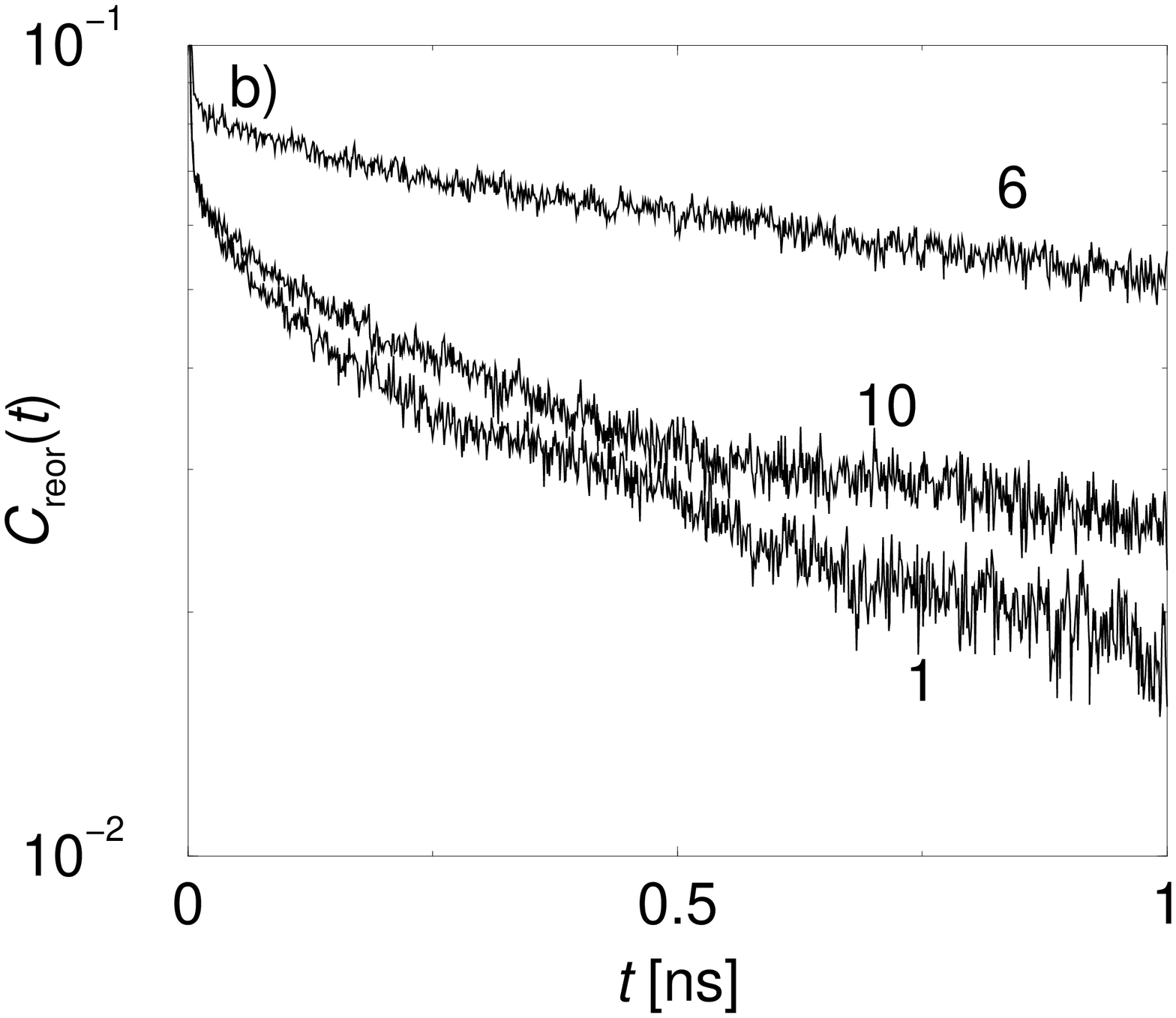}
  \caption{a) Short time reorientation of methyl C$-$H vectors, monomer number
    as in legend (end monomer vs. central monomer,System~3, T=300~K, only
    10-mers are included). b) Long time. Clearly, the two stages are
    separate. A slight difference between the two terminal 
    monomers is visible}
  \label{fig:methyreor}
\end{figure}
In the reorientation of the C$_4-$H vectors of the methyl side group the
two-stage reorientation is clearly visible (Figure~\ref{fig:methyreor}~a). The
first process is non-exponential on the scale of a few
pico-seconds.  The monomer index has almost no influence, only the very local
surroundings contribute. On this time scale, the vector does not experience the
connection to the chain. 

The long time tail, however, is linked to the overall reorientation. The
bonding to the chain leads to a bias in the orientation of the methyl group,
which prevents total decorrelation on the short time scale. The second 
(exponential) process is influenced by chain end effects with decay constants 
of $\tau_{\text{end}}\approx1~\text{ns}$ and
$\tau_{\text{center}}\approx3~\text{ns}$, respectively
(Figure~\ref{fig:methyreor}~b). The decay times were determined by an
exponential fit of the time region between 500 and 1250~ps.

Correlation times for the reorientation of various C$-$H bonds in {\it
cis}-polyisoprene melts were calculated by Moe and Ediger, who used one
long chain under periodic boundary conditions~\cite{moe96a}. 
Table~\ref{tab:cmpmoe} compares the results of this work to their data and the
extreme narrowing limit of the experiments.
\begin{table}
  \[
  \begin{array}{c*{6}{r}*{4}{D{.}{.}{-1}}}
    \hline
    \text{Vector} & 
    \multicolumn{3}{c}{\tau_{\text{reor}}^{trans,\text{sys~1}} [\text{ps}]} &
    \multicolumn{3}{c}{\tau_{\text{reor}}^{trans,\text{sys~2}} [\text{ps}]} &
    \multicolumn{1}{c}{\tau_{\text{reor}}^{trans,A} [\text{ps}]} &
    \multicolumn{1}{c}{\tau_{\text{reor}}^{trans,B}[\text{ps}]} &
    \multicolumn{1}{c}{\tau_{\text{reor}}^{\text{(exp)}} [\text{ps}]} &
    \multicolumn{1}{c}{\tau_{\text{reor}}^{cis} [\text{ps}]}  \\ 
    & \multicolumn{3}{c}{300~\text{K}} 
    & \multicolumn{3}{c}{300~\text{K}} 
    & \multicolumn{1}{c}{413~\text{K}} 
    & \multicolumn{1}{c}{413~\text{K}}  
    & \multicolumn{1}{c}{413~\text{K}}    
    & \multicolumn{1}{c}{413~\text{K}}\\
    \hline
    \opC=\opC     & & 2400 & & & 1900 & & 49   & 51   & -   & -   \\
    \opC_1-\opH & &  810 & & &  640 & & 39   & 25   & 22  & 50  \\
    \opC_2-\opH & & 2400 & & & 1900 & & 50   & 46   & 35  & 75  \\
    \opC_4-\opH & &   -  & & &  480 & &  2.1 &  5.8 & 3.6 &  7.6\\
    \opC_5-\opH & & 1200 & & & 1800 & & 67   & 55   & 26  & 60   \\
    \hline
  \end{array}
  \]
  \caption{Comparison of simulated reorientation times in {\it
      trans}-polyisoprene (this work) and {\it cis}-polyisoprene
    (ref.~\citen{moe96a}) to data determined by extrapolation of
    experiments at lower temperatures into the extreme narrowing limit.
    The {\it cis}-values\cite{moe96a} were determined by numerical
    integration of the first 400~ps. For the {\it trans}-polyisoprene only
    the two innermost monomers are used. At 413~K the analysis was done
    in two ways in order to compare more directly to the Moe and Ediger
    simulations.  $A$: numerical integration and exponential long-time
    tail (see text), $B$: numerical integration to 400~ps. The simulation
    errors are estimated to be about 20\% (difference between systems).
    The reorientation of the methyl group in the $NVT$ simulation could
    not be calculated meaningfully, as there was a force-field problem. (The
    hydrogens were connected to the carbon with an additional torsion,
    which was too strong.)}
  \label{tab:cmpmoe}
\end{table}
At 300~K, the correlation times were determined by numerical integration of
$C_{\text{reor}}(t)$ over the first nano-second and an analytical correction
for the exponential tail. For the methyl groups the numerical integration 
extended only to 20~ps.

Heating to 413~K speeds up the simulation. Here the numerical integration was
performed up to 100~ps, again with an analytical correction, except for the
methyl groups (20~ps) and the C$_1-$H (200~ps) (label
$A$ in Table~\ref{tab:cmpmoe}). For the oligomers, however, the results are
very similar to a numerical integration to 400~ps, which was performed
additionally in order to compare directly to the data by Moe and Ediger (label
$B$ in Table~\ref{tab:cmpmoe}).

The discrepancies between the different systems give a measure of uncertainty
in the results. The reader is reminded that the systems are governed by
slightly differing force-fields and are equilibrated in different manners. 

If the data is compared directly to extrapolated experimental data an overall
discrepancy of about 50\% is found between the {\it trans}-simulations
presented here and the experiments. The integration error in the simulation as
well as the extrapolation of the experiments are sources of error. The systems
are not the same and the simulation model does not reflect reality
perfectly. The experiments themselves are not perfectly reliable. Witt {\it et
al.} showed that NMR experiments for systems as simple as liquid benzene can
result in reorientation times differing by an order of
magnitude~\cite{witt00}.

In the simulation system~2, the hydrogen at C$_5$ reflects too much how the
backbone reorients on the time scale of the double
bond. Experimentally there is a difference between C$_5$-H and C$_2$-H. This
may result from the interaction with the methyl group which repels the H at
C$_5$ very effectively. In system~1 the difference is more pronounced, as the
non-bonded interactions between the methyl hydrogens and the methylene
hydrogen are switched off ($1-5$ interaction, cf. Section~\ref{sec:ff}).
Comparison with experiment does not allow us to decide which scheme for the 
exclusion of nonbonded interactions is more appropriate.

\subsection{Segmental motion}
The double bonds reflect the reorientation of local segments. 
\begin{figure}
  \includegraphics[angle=-90,width=0.5\linewidth]{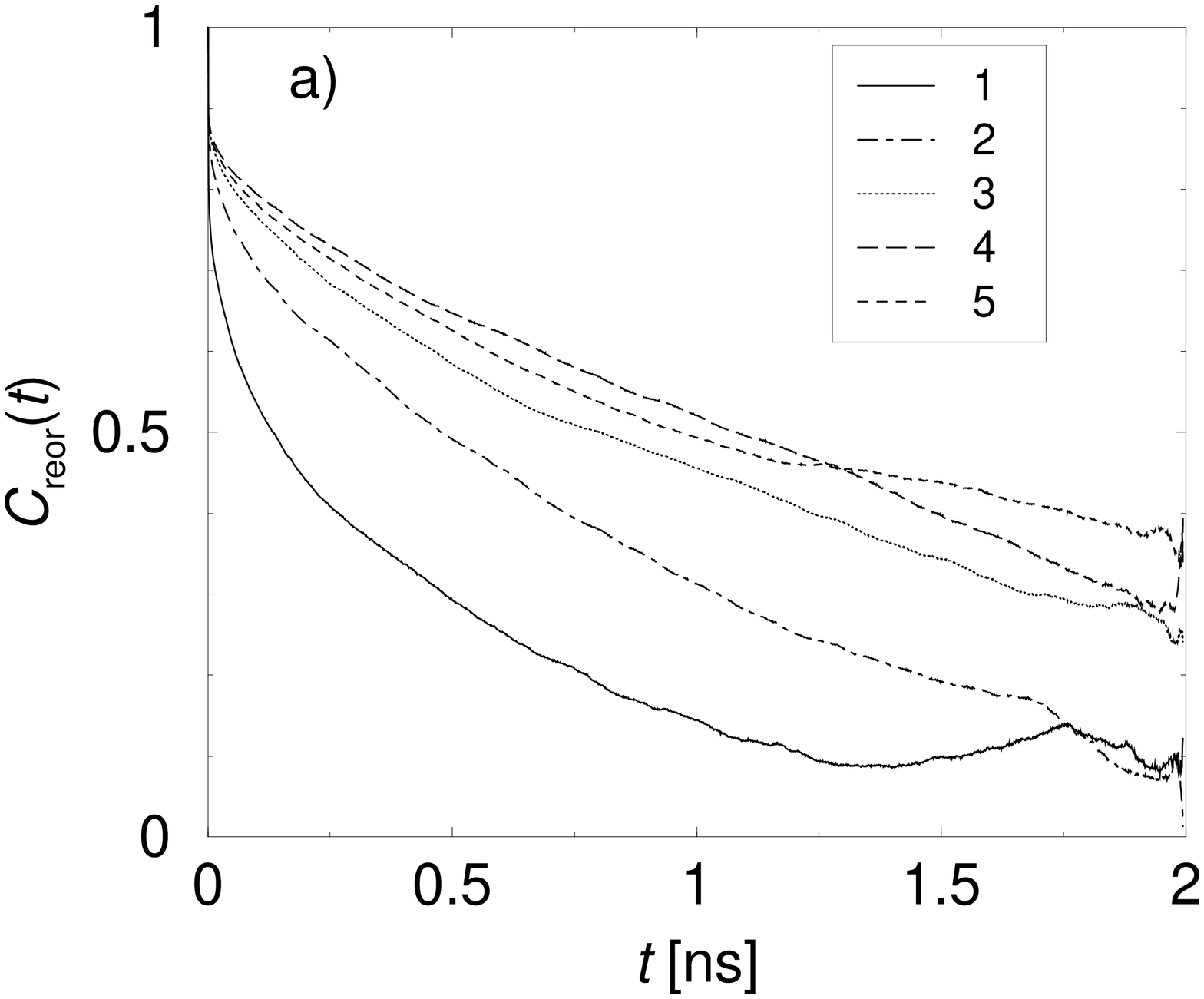}
  \includegraphics[angle=-90,width=0.5\linewidth]{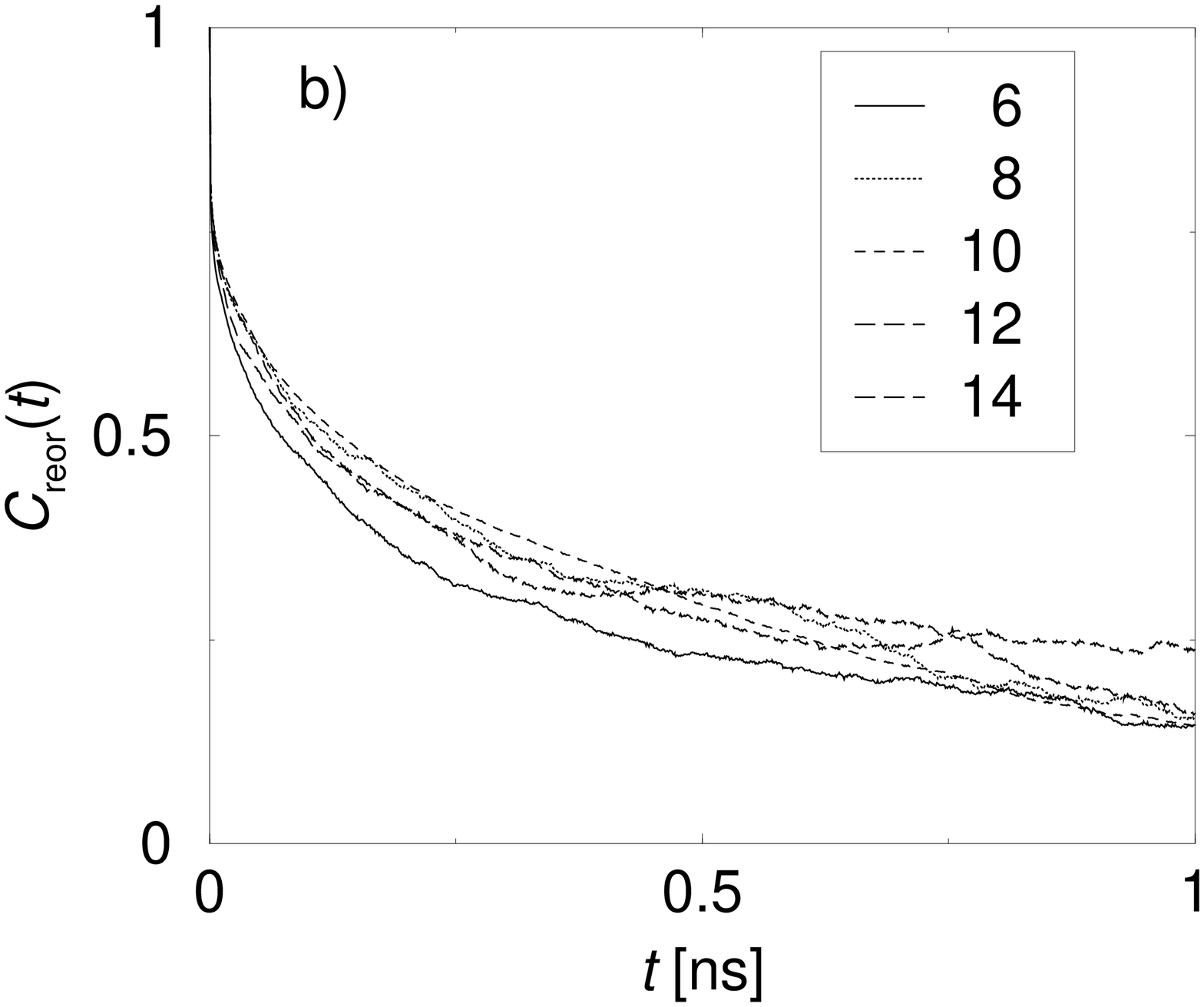}
  \includegraphics[angle=-90,width=0.5\linewidth]{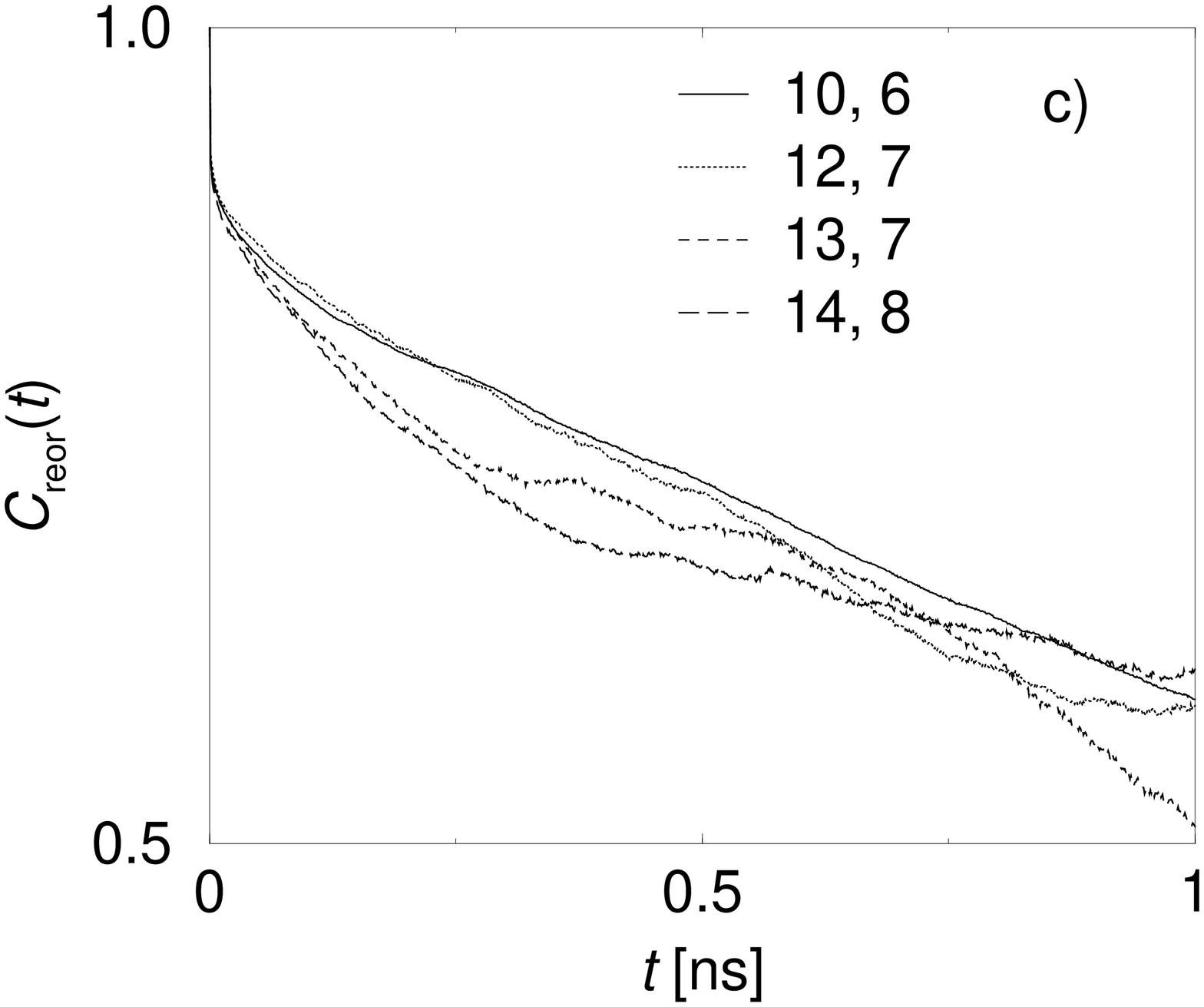}
  \caption{Reorientation of double bonds in system~2 at 300~K: 
    a) Only the chains with ten monomers; results shown for different monomer
    distances from chain end. b) Only the first monomer for different chain
    lengths.  c) Central monomers of the chains of length ten monomers and
    more (the first number denotes the chain length, the second the monomer
    number) in semilogarithmic representation.}
  \label{fig:dblbndreor}
\end{figure}
The ends reorient faster than inner
monomers (Figure~\ref{fig:dblbndreor}a); the six innermost monomers are 
comparable; chains of ten monomer length have a ``bulk'' inner part.

There is a two step reorientation. On very short time scales there is a fast 
drop due to bond angle vibrations. This is not resolved here. For the
inner monomers, a long-time process on the order of the reorientation of the
whole chain (a few nano-seconds) sets in afterwards.

There is little difference in the dynamics of chains of different lengths, at
least in the limited range under study here: the end monomers are free to
move regardless of the rest of the chain (Figure~\ref{fig:dblbndreor}~b) and,
for central monomers, the relaxation is the same within the (large) statistical
error (Figure~\ref{fig:dblbndreor}c). As seen in Figure~\ref{fig:distMW} there
are very few chains of any length other than 10 in system~2 and~3.

\begin{figure}
  \[
  \includegraphics[angle=-90,width=0.5\linewidth]{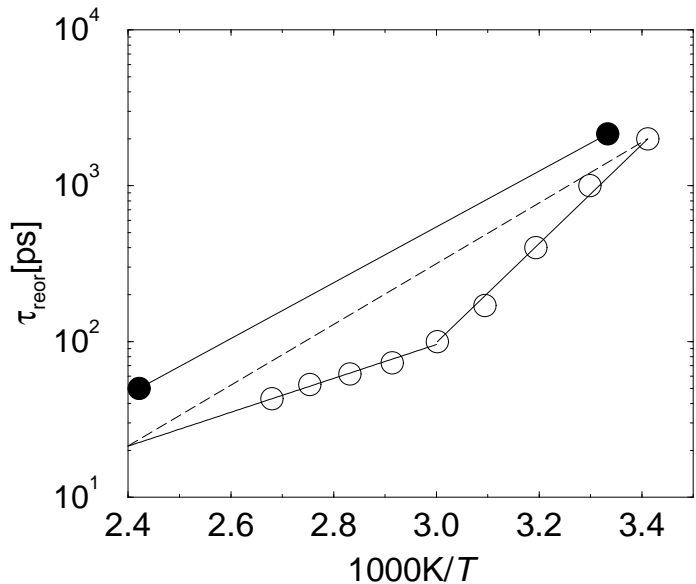}
  \]
  \caption{Arrhenius-plot comparing the C=C reorientation time of
    the simulations (filled circles) to the segmental reorientation time
    inferred by Denault {\it et al.} from their experiments on {\it cis}-PI at
    temperatures between 293~K and 373~K at molecular weights of 7000 to
    130000~\cite{denault89}. For the simulations the values of
    table~\ref{tab:cmpmoe} are averaged at 300~K and 413~K respectively. The
    solid lines are exponential fits to the curves.}
  \label{fig:arrhenius}
\end{figure}

Denault {\it et al.} estimated for chains of molecular weights between
$7\times10^3$g/mol and $1.5\times10^5$~g/mol a segmental reorientation time of
1.0~ns at 303.15~K and of 43~ps at 373.15~K by analyzing their methylene
reorientations~\cite{denault89}. For this, they used the Schaefer
model~\cite{schaefer73} for segmental reorientation, where a
$\chi^2$-distribution of relaxation times is assumed, arising from cooperative
local motion.  The values are of the order of magnitude of the reorientation
data presented here for 300~K. However, the chains are clearly longer and the
focus is on the {\it cis} conformer. The chain length dependence is weak. An
Arrhenius plot of the segmental reorientation time of Denault {\it et. al.} in
comparison to the simulated reorientation of the double bond shows a similar
temperature dependence (Figure~\ref{fig:arrhenius}). The activation energies
deduced are $E_A^{\text{(sim)}}\approx33$~kJ/mol and
$E_A^{\text{(exp)}}=65$~kJ/mol at low temperature and
$E_A^{\text{(exp)}}=19$~kJ/mol at high temperature. The effective experimental
activation energy taking
only the lowest and extrapolated highest point into account arrives at
$E_A^{\text{(exp)}}\approx36$~kJ/mol rather close to the simulation value. To
decide if there is a similar behavior as in experiments with two temperature
regimes more simulations at intermediate temperatures would be necessary.
\section{Mapping onto a simple model}
\label{sec:simple}
As atomistic models such as the one presented here are obviously too demanding
in computer time for long chain and/or long time investigations, mapping onto
simpler models is highly desirable. The simple model we choose here is made of
soft spheres (repulsive Lennard-Jones potential) connected by anharmonic
springs and a weak harmonic bond angle potential leading to a persistence
length of 1.5 monomeric units chosen similar to the atomistic model. For
details of the model and results on structure and dynamics, see
refs.~\citen{faller99b,faller00b}.

\begin{figure}
  \[
  \includegraphics[angle=-90,width=0.5\linewidth]{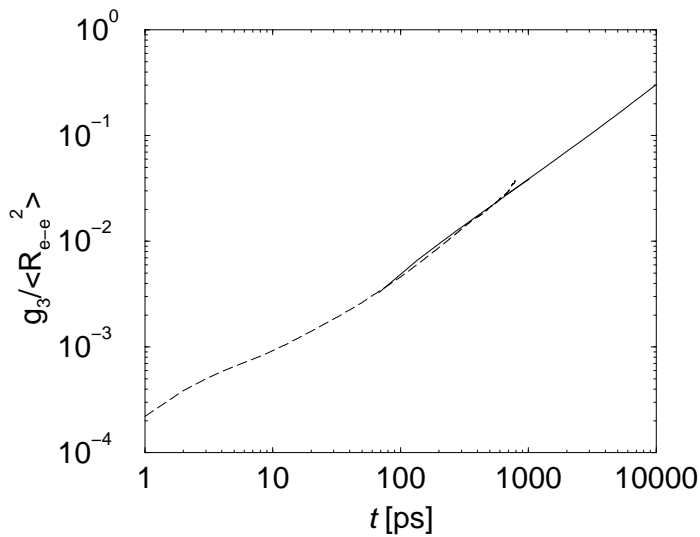}
  \]
  \caption{Mean-square displacements of the center-of-mass in the atomistic
    simulation at 413~K (dashed line) and in the simple model with persistence
    length 1.5 monomer units (solid line). The ordinates are rescaled by the
    respective mean-square end-to-end distances. The time scale of the simple
    model is adjusted to bring the curves into coincidence for a whole order
    of magnitude.}
  \label{fig:map}
\end{figure}

\begin{figure}
  \includegraphics[angle=-90,width=0.5\linewidth]{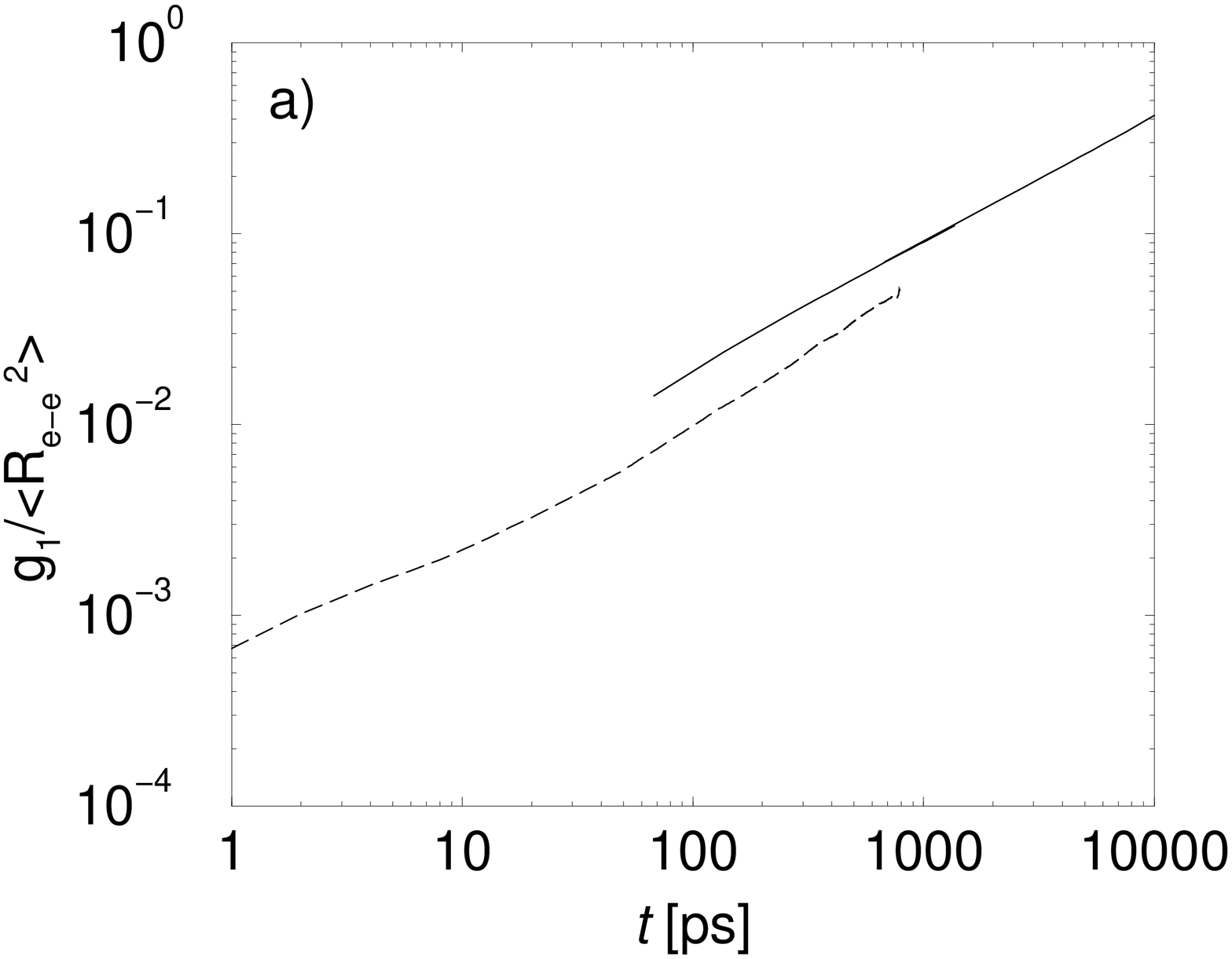}
  \includegraphics[angle=-90,width=0.5\linewidth]{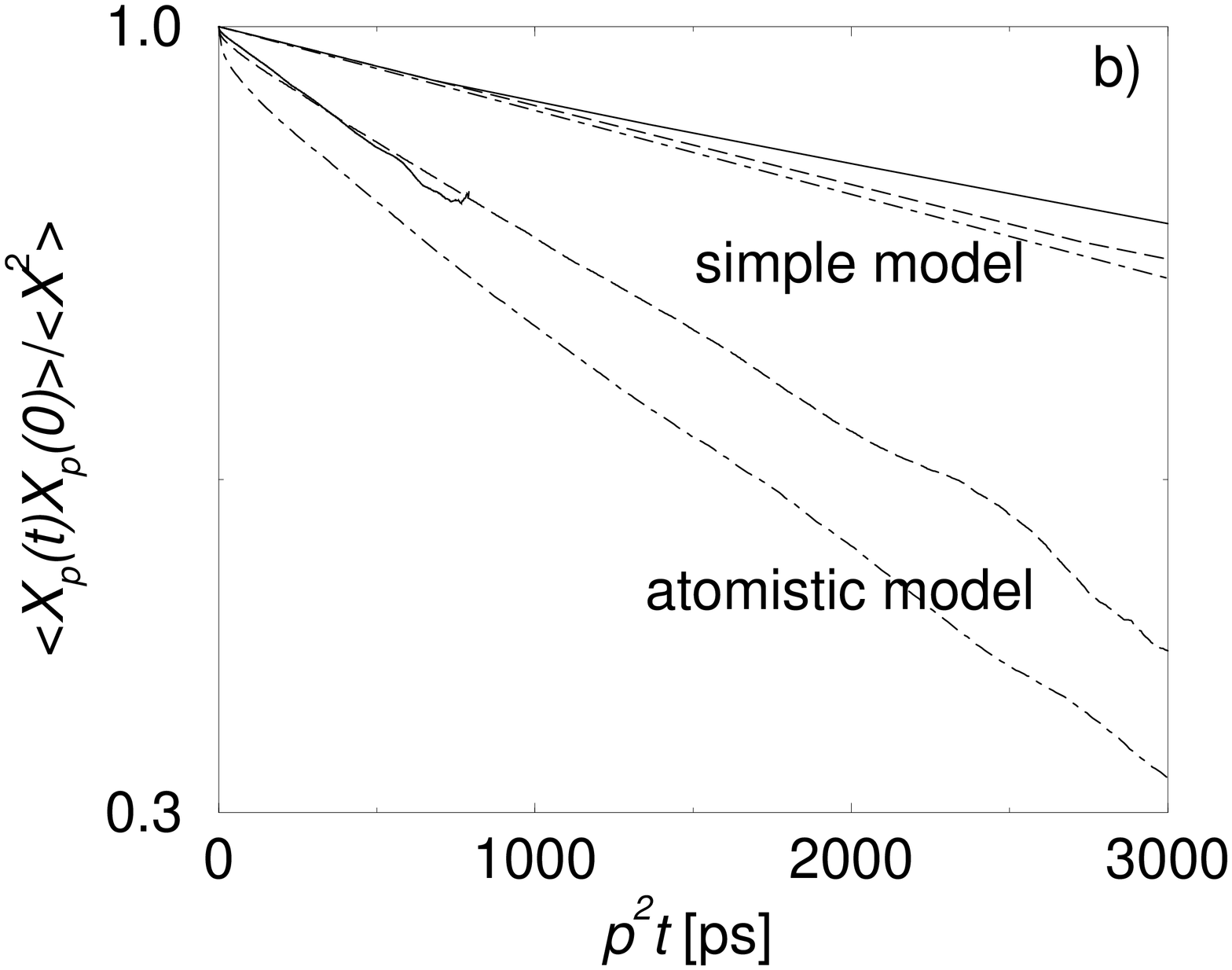}
  \includegraphics[angle=-90,width=0.5\linewidth]{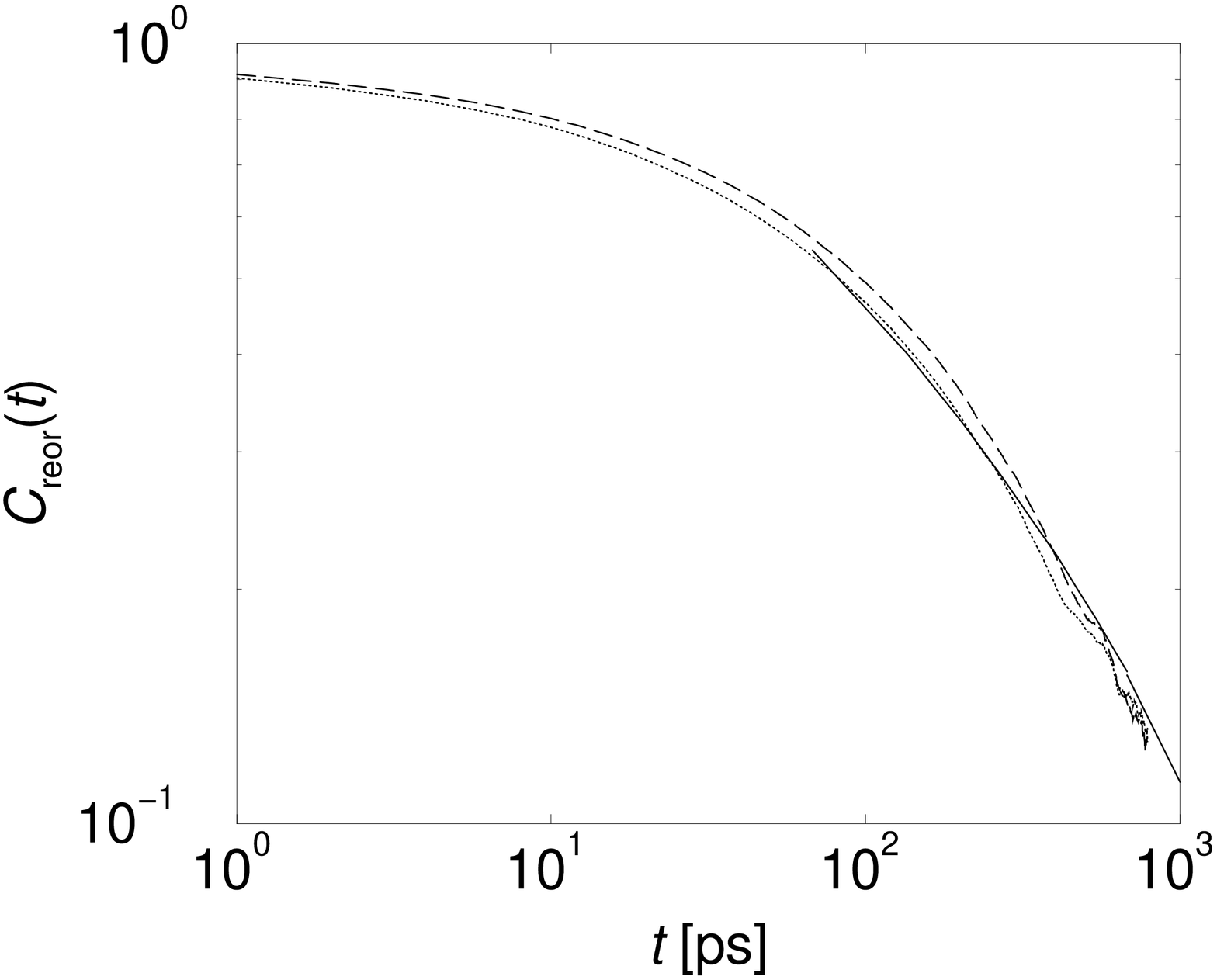}
   \caption{Comparative analysis of the simple model with persistence length
    $l_p=1.5$ and the atomistic simulation at T=413~K. a) Inner monomer
    mean-square displacement (dashed line: atomistic model, center-of-mass of
    double bond of central monomer, solid line: simple model), b) Rouse modes
    (solid lines: first mode, dashed: second mode, dot-dashed: third mode),
    and c) reorientation of nearest neighbor monomer connecting vectors of the
    atomistic model at 413~K compared to the simple model.}
  \label{fig:cmp}
\end{figure}

We have to choose a mapping of one model onto the other before we can
compare. For the mapping of length scales the natural choice is the mean
end-to-end distance. Thus, melts of 10-mers of the atomistic and the simple
models are simulated, then the lengths are set equal. The atomistic 
simulations of system~3b and simulations presented in Ref.~\citen{faller00b} 
are used. The convergence of global chain properties in the atomistic case is 
shown in Figures~\ref{fig:end-end}b and~\ref{fig:map}. For the mapping of time
scales a dynamic quantity is necessary. We map in Figure~\ref{fig:map} the 
center-of-mass mean-square displacements $g_3(t)$ onto each other. Due to 
limited simulation times, a corresponding mapping at 300~K could not be 
accomplished.
With this mapping fixed, comparative analyses of the two models can be
employed in order to check whether both models follow the same
dynamics. Figure~\ref{fig:cmp} shows the respective results for the
mean-square-displacement of inner monomers, $g_1(t)$, the first three Rouse
modes and the reorientation of nearest neighbor monomer-connecting
vectors. These vectors are for the simple model just vectors connecting the
beads. For the detailed model, these are vectors connecting the same carbon in
adjacent monomers, i.e. C$_1-$C$_1$ (dotted line) and C$_2-$C$_2$ (dashed
line). For the Rouse mode analysis of the atomistic model the centers-of-mass
of the double bond are taken as monomer positions.
The first three Rouse modes and the inner-monomer-MSD do not coincide
perfectly within this mapping. Still, they differ by only about a factor of
two to three. The reorientation of the monomer-monomer vector even coincides
without any refinement. The two differently defined vectors in the atomistic
models are almost indistinguishable. Thus, on this scale the atomistic
details are already rather unimportant; they may be incorporated into an
intrinsic chain stiffness. These results show that the local dynamics are not
exactly the same for the atomistic and the simple (coarse-grained) model.
There are striking similarities, however, which are much stronger than one
might expect as the models are completely different. This finding confirms the
concept that the atomistic details on the scale of more than a monomer play
only the role of shaping a persistence length. With the simple model we have
carried out several 
investigations\cite{faller99b,faller99d,faller00b,faller00sa}, especially of
local reorientation. 

One might argue, that discrepancies of a factor of three are no small effects. 
However, keeping in mind that the two models are drastically different one 
would expect on first sight a much stronger discrepancy. We performed the 
extreme step from the smallest possible model without quantum effects to a
model where the complete identity of the polymer is only regarded by means of 
its stiffness.The successful mapping to the atomistic model is a further 
(retrospective) validation of those results. A more detailed investigation of 
the mapping will be shown in a separate publication.~\cite{faller00sc}
\section{Conclusions}
This simulation study is concerned with decamers of {\it trans}-polyisoprene,
which are short in comparison to experimental systems. Yet, the simulations
are able to describe correctly local structural and dynamical features of the
polymer. This can be seen by comparing results like chain extension, structure
functions and correlation times for C$-$H vectors obtained by NMR. Taking into
account that the simulated and experimental systems are not always identical
in composition, temperature and so on, the agreement is quite good. We have
also recently studied the free volume properties of our melts and compared
them to positronium annihilation data\cite{schmitz00s}, and the agreement is
again very good. We thus conclude that our all-atom model provides a faithful
description of this polymer.

With the model, we analyzed the packing of chains. The detailed analysis of
several inter-chain radial distribution functions shows that monomers approach
each other most strongly with the exposed methyl groups followed by the
methylene groups, whereas the vinyl carbons tend to be less accessible. All
specificity in the interaction is however limited to the first solution
shell. This is also seen in the mutual orientation of tangent vectors of
neighboring chains. Chains have the tendency to be parallel at the first
neighbor distance. The orientational correlation between second neighbors is
already very small. From comparing different choices for the tangent vectors
it is also evident that the orientations of bond vectors (i.e. short vectors)
show some structure arising from atomistic interactions, whereas the
orientation between inter-monomer (long) vectors is less structured and
already close to what is found for a generic bead-spring
model\cite{faller99d,faller99b}. 

The reorientation dynamics of bond (C$-$C and C$-$H) vectors typically follows
a two-stage process, a fast (picosecond) relaxation due to local vibrations
followed by a long-time reorientation characteristic for the reorientation of
the parent polymer segment. The relative contribution of both processes to the
overall reorientation is in good agreement with estimates from NMR
measurements. 

In order to study the non-local structure and dynamics of {\it
  trans}-polyisoprene extensive simulations have been performed with a generic
bead-spring model which was only augmented by an intrinsic bending
stiffness\cite{faller00b,faller00sa}. We have successfully mapped this model
onto the present atomistic simulation. After matching length and time scales,
all characteristic time constants then agree to within a factor of $2-3$. This
illustrates that it is possible to develop polymer models at different levels
of detail and have a description of polymer dynamics which smoothly connects
both time scales. Systematic protocols for mapping atomistic to coarse-grained
models and back are therefore being developed in our laboratory also for other
polymer systems\cite{tschoep98a,tschoep98b,meyer00,reith00s}.

An interesting technical point is the comparison between the polymer samples
equilibrated with end-bridging Monte Carlo (EBMC) and those without. The EBMC 
proves to be a useful tool in the initial equilibration of polymer melts as it 
offers a way to wander through phase space more efficiently. At a global 
structural level, the EBMC provides better equilibrated samples. However, it 
is quite surprising that strictly local properties like monomer packing or
bond-vector reorientation do not seem to be affected appreciably by the way
the sample is prepared.
\section*{Acknowledgments}
We'd like to thank Kurt Kremer and Heiko Schmitz for fruitful
discussions. Financial support from the German ministry of research (BMBF) as
well as the TMR program of the European Union is gratefully acknowledged.
\bibliography{standard}
\bibliographystyle{macromolecules}
\end{document}